\documentclass[aps,prb,twocolumn,groupedaddress]{revtex4}

\usepackage{graphicx}
\usepackage{amssymb}
\usepackage{amsmath}
\usepackage{amsfonts}
\usepackage{amssymb}
\usepackage{graphicx}
\usepackage{hyperref}
\usepackage{color}

\newcommand{\ket}[1]{|{#1}\rangle}
\newcommand{\bra}[1]{\langle{#1}|}
\newcommand{\boldrho}{\mbox{\boldmath$\rho$}}
\newcommand{\rs}{\rm \scriptscriptstyle}

\begin{document}

\title{%Interaction designing:\\
 Condensed Matter Physics with Cold Polar Molecules}

\author{G. Pupillo$^{1,2}$, A. Micheli$^{1,2}$, H.P. B\"uchler$^{3}$, and P. Zoller$^{1,2}$}
\affiliation{$^{1}$Institute for Theoretical Physics, University of
Innsbruck,  6020 Innsbruck, Austria} \affiliation{$^{2}$Institute
for Quantum Optics and Quantum Information, 6020 Innsbruck, Austria}
\affiliation{$^{3}$Institute for Theoretical Physics III, University
of Stuttgart, Pfaffenwaldring 57, 70550 Stuttgart, Germany}

\date{\today}

\begin{abstract}
\end{abstract}

%\pacs{03.75.Fi, 05.30.Jp, 32.80.Pj}

\maketitle

\section{Introduction}The realization of Bose Einstein condensates and quantum degenerate
Fermi gases with cold atoms has been one of the highlights of
experimental atomic physics during the last decade~\cite{Ultracold},
and in view of recent progress in preparing cold molecules we expect
a similarly spectacular development for molecular
ensembles~\cite{SpecialIssue1,Ex1,Bethlem00,Ex04,ExpYeJin,ExpGrimm,ExpDen,Sawyer07,Stwalley04,Crompvoets01,Rempe04,Hinds04,Ex2,Wang,Ex4,Ex5,Hudson31,EX99,DeMille02,Greiner03,Regal03,EX03}.
The outstanding features of the physics of cold atomic and molecular
gases are the microscopic knowledge of the many-body Hamiltonians,
as realized in the experiments, combined with the possibility to
control and tune system parameters via external fields. Examples are
the trapping of atoms and molecules with magnetic, electric and
optical traps, allowing for the formation of quantum gases in 1D, 2D
and 3D geometries, and the tuning of {\em contact} inter-particle
interactions by varying the scattering length via Feshbach
resonances~\cite{Fano61,Duine04}. This control is the key for the
experimental realization of fundamental quantum phases, as
illustrated by the BEC-BCS crossover in atomic Fermi
gases~\cite{Regal04a,Bartenstein04,Zwierlein05,Partridge05,Chin04},
the Kosterlitz-Thouless transition~\cite{Dalib06} and the
superfluid-Mott insulator quantum phase transition with cold bosonic
atoms in an optical lattice~\cite{Bloch02,Spielman06}. A recent highlight has been the
realization of a degenerate magnetic dipolar gas of $^{52}$Cr
atoms~\cite{Pfau1,Pfau2,Pfau3}. Below we will be mainly interested in heteronuclear molecules prepared in their electronic and vibrational ground state.  The new feature of polar molecules is their  \emph{large electric dipole moments} associated with rotational excitations. The new aspects in a condensed matter theory of cold polar molecules are the \emph{large dipole-dipole interactions} between molecules. This points towards the possibility of manipulating these strong and long-range interactions with external DC and AC microwave fields. In particular, this raises interesting questions of cold ensembles of polar molecules as \emph{strongly correlated systems}\cite{all0,all1,all2,all3,all4,all5,all6,all7,all8,all9,all10,all11,all12,Buechler07,Astrakharchik07,Mora07,Ark1,all13,all14,micheli06,brennenNJP07,Pupillo08,excitons}.

Magnetic or electric
dipole moments in external fields can give rise to anisotropic,
long-range dipole-dipole interactions. Analogous to the case of
cold atoms with contact interactions, a key element for the
realization of interesting quantum phases and phase transitions with
interacting dipolar gases is the capability of controlling and
tuning system parameters using external fields. Much work has been
recently devoted to the study of cold collisions in
dipolar gases~\cite{col1,col2,col3,col4,col5,col6,Micheli07}, which in this book is
reviewed in the chapters contributed by Hutson, Bohn, and Dalgarno. In the context of degenerate molecular gases a significant body of recent work has focused on
 the regime of {\em weak  interactions},
where the isotropic contact interaction potential competes with the
anisotropic long range dipole-dipole interaction. For example, the existence of
rotons in weakly-interacting dipolar gases has been
predicted~\cite{Rot1,Rot3,Rot4,Rot5,Rot6,Rot7,Rot8,Rot9,Rot10},
while exciting prospects have been envisioned for rotating
systems~\cite{Rota1,Rota2,Rota3,Rota4,Rota5,Rota6} and polar molecules in
optical
lattices~\cite{Opt1,Opt2,Opt3,Opt4,Opt5,Opt6,Opt7,BuchlerNature07}.
Below we will be mainly interested in the many-body dynamics of polar molecules in the {\em
strongly interacting limit}. In particular, we will discuss a toolbox for engineering interesting many-body Hamiltonians based on the manipulation of the electric dipole moments with external DC and AC fields, and thus of the molecular interactions. This forms the basis for the realization of novel quantum phases in these systems. Our emphasis will be on condensed matter aspects, while we refer to the contribution by Yelin, DeMille and Cote in the present book for application in the context of quantum information processing.

This review is organized as follows. In Sect.~\ref{sec:secIntro2} we give a qualitative tour through some of the key ideas of engineering Hamiltonians and of the associated quantum phases.  This is followed by two slightly
more technical sections, Sect.~\ref{sec:secMolecularHam} and
Sect.~\ref{Sec:TwoMolecules}, where we provide some details of the
realization of a 2D setup where particles interact via purely
repulsive $1/r^3$ potentials, and where we sketch how to design more
complicated interactions by using a combination of AC and DC fields.
Finally, Sect.~\ref{sec:secApplications} deals with the applications of the engineering
of interaction potentials in the context of the realization of
strongly correlated phases and quantum simulations.

\section{Overview: Strongly Interacting Systems of Cold Polar Molecules}

\label{sec:secIntro2}

In this section we give a qualitative overview of many-body physics
of cold polar molecules with emphasis on strongly interacting systems.
In the following sections we will return to the various topics in
a more in-depth discussion.

\subsection{Effective many body Hamiltonians}
\begin{figure*}
  \begin{center}
    \includegraphics[width=\textwidth]{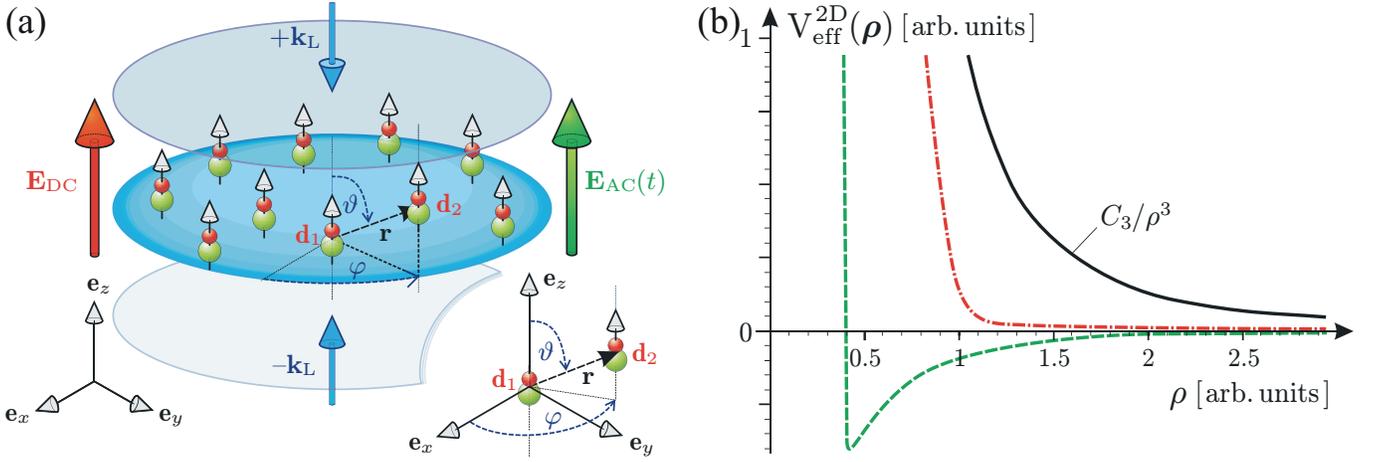}%[width=\columnwidth]{figs/Figure_1_copy.eps}
  \end{center}
  \caption{\label{fig:fig1}(a) System setup: Polar molecules are
    trapped in the ($x,y$)-plane by an optical lattice made of two
    counter-propagating laser beams with wavevectors $\pm{\bf k}_{\bf
      L}=\pm{k}_{\bf L}{\bf e}_z$, (blue arrows). The dipoles are
    aligned in the $z$-direction by a DC electric field ${\bf E}_{\rm
      DC}\equiv E_{\rm DC}{\bf e}_z$ (red arrow). An AC microwave
    field is indicated (green arrow).
    Inset: Definition of polar ($\vartheta$) and azimuthal
      ($\varphi$) angles for the relative orientation of the
      inter-molecular collision axis ${\bf r}_{12}$ with respect to a
      space-fixed frame, with axis along $z$. (b) Qualitative sketch of effective 2D
  potentials
  $V_{\rm eff}^{\rm 2D}(\boldrho)$ for polar
  molecules confined in a 2D (pancake) geometry. Here,
  $\boldrho=r_{12} \sin\vartheta(\cos\varphi,\sin\varphi,0)$ is the 2D coordinate
  in the plane $z=0$ and $\rho=r_{12} \sin\vartheta$ (see Fig.~\ref{fig:fig1}, inset).
   Solid line: Repulsive dipolar potential $V_{\rm eff}^{\rm 2D}(\boldrho)=D/\rho^3$ induced by a DC electric field. Dash-dotted line: ``Step-like'' potential
  induced by a single AC microwave field and a weak DC field. Dashed
  line: Attractive potential induced by the
  combination of several AC fields and a weak DC field.
  The potentials
  $V_{\rm eff}^{\rm 2D}(\boldrho)$ and the separation $\rho$ are given in arbitrary units.}
\end{figure*}

Hamiltonians underlying condensed matter physics of $N$ structureless
bosonic or fermionic particles have the generic form

\begin{equation}
H_{\textrm{eff}}=\sum_{i=1}^{N}\left[\frac{\mathbf{p}_{i}^{2}}{2m}+V_{\mathrm{trap}}(\mathbf{r}_{i})\right]+V_{\textrm{eff}}^{{\rm
3D}}\left(\{{\bf
r}_{i}\}\right),\label{eq:NbodyHamiltonian}\end{equation}
where $\mathbf{p}_{i}^{2}/2m$ is the kinetic energy term, and
$V_{\mathrm{trap}}(\mathbf{r}_{i})$
is a confining potential for the particles. The term $V_{\rs
eff}^{{\rm 3D}}\left(\{{\bf r}_{i}\}\right)$
represents an effective $N$-body interaction, which can be expanded
as a sum of two-body, three-body interactions \emph{etc.,} \begin{equation}
V_{\rs eff}^{{\rm 3D}}\left(\{{\bf
r}_{i}\}\right)=\sum_{i<j}^{N}V^{{\rm 3D}}\left({\bf r}_{i}\!-\!{\bf
r}_{j}\right)+\!\sum_{i<j<k}^{N}W^{{\rm 3D}}\left({\bf r}_{i},{\bf
r}_{j},{\bf r}_{k}\right)+\ldots,\label{effint1}\end{equation}
where in most cases only two-body interactions are considered. Typically,
$V_{\rs eff}^{{\rm 3D}}\left(\{{\bf r}_{i}\}\right)$ must be interpreted
as \emph{effective} interactions valid in a \emph{low energy theory},
obtained by integrating out the high energy degrees of freedom of
the system. In the following we will discuss the derivation of such
many body Hamiltonians for a cold ensemble of polar molecules in the
electronic and vibrational ground state, and show how the effective
interactions in these systems can be ``designed'' on an $N$-body
level via control of rotational excitations with external fields,
in a form which is unique to polar molecules.

Our starting point is the Hamiltonian for a gas of cold heteronuclear
molecules prepared in their electronic and vibrational ground-state,
\begin{eqnarray}
H(t) & = & \sum_{i}^{N}\left[\frac{\mathbf{p}_{i}^{2}}{2m}+V_{\mathrm{trap}}(\mathbf{r}_{i})+H_{\mathrm{in}}^{(i)}-\mathbf{d}_{i}\mathbf{E}(t)\right]\nonumber
\\
 & + & \sum_{i<j}^{N}V_{{\rm
dd}}(\mathbf{r}_{i}-\mathbf{r}_{j}).\label{eq:eqGeneral}\end{eqnarray}
Here the first term in the single particle Hamiltonian corresponds
to the kinetic energies of the molecules, while
$V_{\mathrm{trap}}(\mathbf{r}_{i})$
represents a possible trapping potential, as provided, for example,
by an optical lattice, or an electric or magnetic traps. The term
$H_{{\rm in}}^{(i)}$ describes the internal low energy excitations
of the molecule, which for a molecule with a closed electronic shell
${}^{1}\Sigma(\nu=0)$ (e.g. of the type SrO, RbCs or LiCs), correspond
to the rotational degree of freedom of the molecular axis. This term
is well described by a rigid rotor $H_{{\rm in}}^{(i)}\equiv
H_{\mathrm{rot}}^{(i)}=B\mathbf{J}_{i}^{2}$
with $B$ the rotational constant (in the few to tens of GHz regime)
and $\mathbf{J}_{i}$ the dimensionless angular momentum. The rotational
eigenstates $|J,M\rangle$ for a quantization axis $z$, and with
eigenenergies $BJ(J+1)$ can coupled by a static (DC) or microwave
(AC) field $\mathbf{E}$ via the \emph{electric} dipole moment $\mathbf{d}_{i}$,
which is typically of the order of a few Debye. For
distances outside of the molecular core, the two-body interaction
as given by the second line of Eq.~\eqref{eq:eqGeneral} is the dipole-dipole interaction
\begin{eqnarray}
V_{{\rm dd}}({\bf r})=\frac{{\bf d}_{i}\cdot{\bf d}_{j}-3\left({\bf
d}_{i}\cdot{\bf e}_{r}\right)\left({\bf e}_{r}\cdot{\bf
d}_{j}\right)}{r^{3}}.\label{eq:eqDipDip1}\end{eqnarray}
Here $r\equiv|{\bf r}|=|{\bf r}_{i}-{\bf r}_{j}|$ denotes the distance
between two polar molecules, with ${\bf e}_{r}$ the unit vector along
the collision axis. In view of the large electric dipole moments this
term provides a comparatively strong, long-range anisotropic interaction
between the molecules.

The many body dynamics of cold polar molecules is thus governed by
an interplay between dressing and manipulating the rotational states
with DC and AC fields, and strong dipole-dipole interactions. In the absence of electric fields, the molecules prepared in a ground rotational
state $J=0$ have no net dipole moment, and interact via a van-der-Waals
attraction $V_{\mathrm{vdW}}\sim-C_{6}/r^{6}$, reminiscent of the interactions
of cold Alkali atoms in their electronic ground-states. Electric fields
admix excited rotational states and induce static or oscillating dipoles,
which will interact via strong dipole-dipole interactions $V_{dd}$
with the characteristic $1/r^{3}$ dependence of Eq.~\eqref{eq:eqDipDip1}.
Note that two parallel dipoles repel each other, while dipoles aligned
along the collision axis will attract each other, possibly inducing
instabilities in a many-body system. Thus to obtain stable many-particle
phases is often possible only in reduced geometries, i.e. in combination
with an external trapping potential $V_{\mathrm{trap}}(\mathbf{r}_{i})$. Finally, we emphasize that microwave excited rotational states of polar molecules are long-lived, which makes these states available without the penalty of introducing significant decoherence. This is in contrast to atomic systems, where spontaneous emission from laser excited electronic states is one of the main contributions to decoherence.

The connection between the full molecular $N$-particle Hamiltonian
(\ref{eq:eqGeneral}) including rotational excitations and dressing
fields, and the effective Hamiltonian (\ref{eq:NbodyHamiltonian})
can be made in a Born-Oppenheimer approximation. If we diagonalize
for frozen spatial positions $\{{\bf r}_{i}\}$ of the $N$ molecules
the Hamiltonian~\cite{Note1}
$H_{BO}=\sum_{i}^{N}\left[H_{\mathrm{in}}^{(i)}-\mathbf{d}_{i}\mathbf{E}\right]+\sum_{i<j}^{N}V_{{\rm
dd}}(\mathbf{r}_{i}-\mathbf{r}_{j}),$
we obtain a set of energy eigenvalues $V_{\textrm{eff}}^{{\rm
3D}}\left(\{{\bf r}_{i}\}\right)$,
which are interpreted as the effective $N$-particle potential in
the single channel many-body Hamiltonian (\ref{eq:NbodyHamiltonian}).
The dependence of $V_{\textrm{eff}}^{{\rm 3D}}\left(\{{\bf r}_{i}\}\right)$
on the electric fields $\mathbf{E}$ provides the basis of the engineering
of the many body interactions of two-body, three-body terms \emph{etc.}
in (\ref{effint1}). The validity of this adiabatic approximation
and of the associated decoupling of the Born-Oppenheimer channels will be
discussed below.

The above considerations set the stage for a discussion of engineering
many-body Hamiltonians for polar molecules, and associated quantum
phases. In the following subsections we will discuss specific examples
of DC and AC field configurations for designing two-body and three-body
interactions, which will be the content of Sects.~\ref{sec:secSelfAssembled} and~\ref{sec:secBlueThree} below. Our discussion
can also be adapted to optical lattices, and thus to a derivation
of Hubbard models for polar molecules (Sect.~\ref{sec:secHubLatt}). Furthermore, we
can extend these derivations to molecules with internal degrees of
freedom representing a spin (Sect.~\ref{sec:secLattSpin}). Extended Hubbard models with couplings to phonons are presented in Sect.~\ref{sec:secFloat} for molecules trapped in self-assembled dipolar cystals.

\subsection{Self-assembled crystals.}\label{sec:secSelfAssembled}

\begin{figure*}%[t!]%[htbp!]
\begin{centering}
\includegraphics[width=\textwidth]{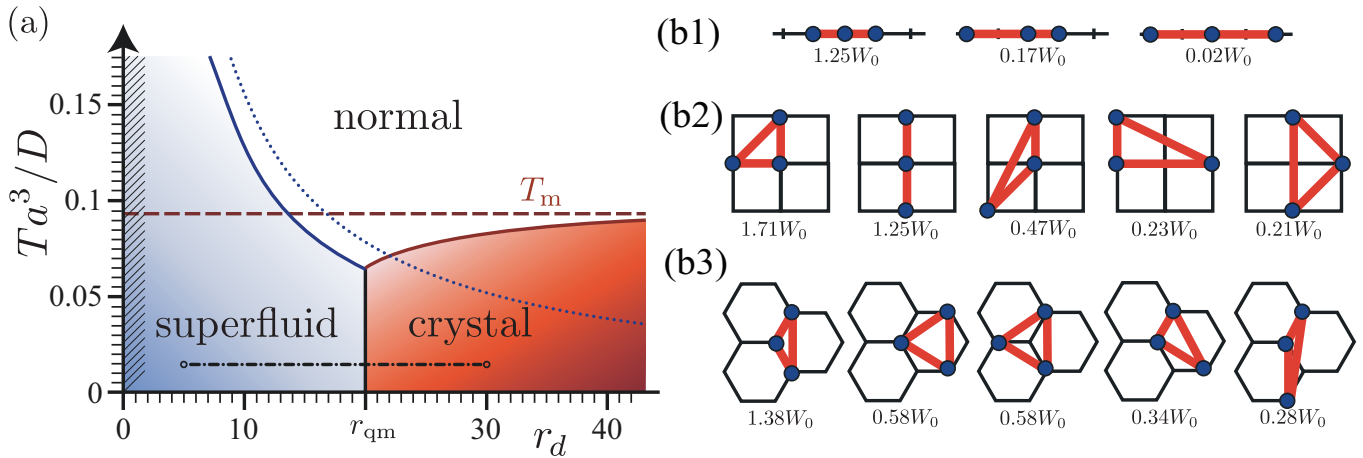}
\par\end{centering}
\caption{\label{figs:fig1000}(a) Tentative phase diagram in the $T-r_{d}$ plane:
crystalline phase for interactions
$r_{d}>r_{\mathrm{\scriptscriptstyle QM}}$ and temperatures below
the classical melting temperature $T_{m}$ (dashed line). The
superfluid phase appears below the upper bound $T<\protect\pi
\hbar^{2}n/2 m $ (dotted line). The quantum melting transition is
studied at fixed temperature $T=0.014 D/a^{3}$ with
interactions $r_{d}=5-30$ (dash-dotted line). The crossover to the
unstable regime for small replusion and finite confinement
$\omega_{\perp}$ is indicated (hatched region).
(b) Strengths of the
dominant three-body interactions $W_{i j k}$ appearing in the
Hubbard model of Eq.~\eqref{Hubbard} for different lattice geometries: (b1)
one-dimensional setup; (b2) two-dimensional square lattice; (b3)
two-dimensional honeycomb lattice.  The characteristic energy scale
$W_{0}= \gamma_{2} D R_{0}^6/a^6$ is discussed following
Eq.~(\ref{threebodystrength}).}
\end{figure*}
The conceptually simplest example, although remarkably
rich from a physics point of view, is a system of cold polar molecules in a DC electric
field under conditions of strong transverse confinement. The setup
is illustrated in Fig.~\ref{fig:fig1}(a). A weak DC field along
the $z$-direction induces a dipole moment $d$ in the ground state
of each molecule. These groundstate molecules interact via the effective
dipole-dipole interaction $V_{{\rm eff}}^{{\rm 3D}}({\bf r})=D(r^{2}-3z^{2})/r^{5}$
according to their induced dipoles, with $D=d^{2}$. For molecules confined
to the $x,y$-plane perpendicular to the electric field this interaction
is purely repulsive. For molecules displaced by $z>r/\sqrt{3}$ the interaction
becomes attractive, indicating an instability in the many body system.
This instability is suppressed by a sufficiently strong 2D confinement
with potential a $V_{{\rm trap}}(z_{i})$ along $z$, due to, for
example, an optical potential induced by an off-resonant light field~\cite{Buechler07}.

The 2D dynamics in this pancake configuration is described by the
Hamiltonian \begin{equation}
H_{{\rm eff}}^{{\rm 2D}}=\sum_{i}\frac{{\bf p}_{\mathbf{\rho}i}^{2}}{2m}+\sum_{i<j}V_{{\rm eff}}^{{\rm 2D}}(\boldrho_{ij}),\label{hamilton1}\end{equation}
which is obtained by integrating out the fast $z$-motion. Equation~(\ref{hamilton1})
is the sum of the 2D kinetic energy in the $x$,$y$-plane and a repulsive
2D dipolar interaction \begin{eqnarray}
V_{{\rm eff}}^{{\rm 2D}}(\boldrho)=D/\rho^{3},\label{eq:eqIn}\end{eqnarray}
with $\boldrho_{ij}\equiv(x_{j}-x_{i},y_{j}-y_{i})$ a vector in the
$x,y$-plane {[}solid line in Fig.~\ref{fig:fig1}(b)]. The distinguishing
feature of the system described by the Hamiltonian (\ref{hamilton1})
is that tuning the induced dipole moment $d$ drives the system
from a weakly interacting gas (a 2D superfluid in the case of bosons),
to a crystalline phase in the limit of strong repulsive dipole-dipole
interactions. This transition and the crystalline phase have no analog
in the familiar atomic bose gases with short range interactions modelled
by a pseudopotential of a given scattering length.

A crystalline phase corresponds to the limit of strong repulsion where
particles undergo small oscillations around their equilibrium positions,
which is a result of the balance between the repulsive long-range
dipole-dipole forces and an additional (weak) confining potential
in the $x,y$-plane. The relevant parameter is\begin{eqnarray}
r_{d}\equiv\frac{E_{\textrm{pot}}}{E_{\textrm{kin}}}=\frac{D/a^{3}}{\hbar^{2}/ma^{2}}=\frac{Dm}{\hbar^{2}a},\label{eq:eqrd}\end{eqnarray}
which is the ratio of the the interaction energy and the kinetic energy
at the mean interparticle distance $a$. This parameter is tunable
as a function of $d$ from $r_{d}$ small to large. A crystal will
form for $r_{d}\gg1$, when interactions dominate. For a dipolar crystal, this is the limit of large densities,
where typically collisions become harmful. However the crystalline phase
will protect a cold ensemble of polar molecules from (harmful) close-encounter
collisions. This density dependence is in contrast to Wigner crystals
with $1/r$- Coulomb interactions, as realized e.g. with laser cooled
trapped ions~\cite{ions}. In this case $r_{c}=(e^{2}/a)/\hbar^{2}/ma^{2}\sim a$
and the crystal forms at low densities. In addition, the charge $e$
is a fixed quantitiy, while $d$ can be varied as a function of the
DC field.

Fig.~\ref{figs:fig1000}(a) shows a tentative phase diagram for a
dipolar gas of bosonic molecules in 2D as a function of $r_{d}$ and
temperature $T$. In the limit of weak interactions $r_{d}<1$, the
ground state is a superfluid (SF) representing a finite (quasi-)condensate.
%The SF is characterized by a superfluid fraction $\rho_{s}(T)$, which
%depends on temperature $T$, with $\rho_{s}(T=0)=1$. A Berezinskii--Kosterlitz--Thouless
%transition~\cite{BKT72,BKT73} towards a normal fluid is expected to occur at a finite
%temperature $T_{\mathrm{{\scriptscriptstyle KT}}}=\pi\rho_{s}\hbar^{2}n/2m$.
In the opposite limit of strong interactions $r_{d}\gg1$ the polar
molecules are in a crystalline phase for temperatures $T<T_{m}$ with
$T_{m}\approx0.09D/a^{3}$, see Ref.~\cite{Kalia}. The configuration with
minimal energy is a triangular lattice with excitations given by acoustic phonons. In Ref.\cite{Buechler07}
we investigated the intermediate strongly interacting regime with
$r_{d}\gtrsim1$, using a recently developed Path-Integral Monte-Carlo
technique (PIMC)~\cite{Boninsegni06}, and we determined the critical
interaction strength $r_{\mathrm{{\scriptscriptstyle QM}}}$ for the
quantum melting transition from the crystal into the superfluid. We
found the latter to occur at $r_{\mathrm{{\scriptscriptstyle QM}}}=18\pm4,$
a result which has been confirmed with a number of quantum Monte-Carlo
techniques~\citet{Astrakharchik07,Mora07}.

In Sect.~\ref{sec:secApplications} below we return to a more detailed
discussion of these quantum phases, and in particular of the crystalline phase,
and we show that the relevant parameter regime where these phases occur is accessible with
polar molecules. Besides the fundamental interest in dipolar quantum
gases, the crystalline phase has interesting applications, e.g. in
the context of quantum information~\cite{all14}. We will return to self-assembled
dipolar lattices below in a discussion of Hubbard models.

\subsection{Blue-shielding and three-body interactions}\label{sec:secBlueThree}
\begin{figure*}%[htb]
\begin{center}
\includegraphics[width=\textwidth]{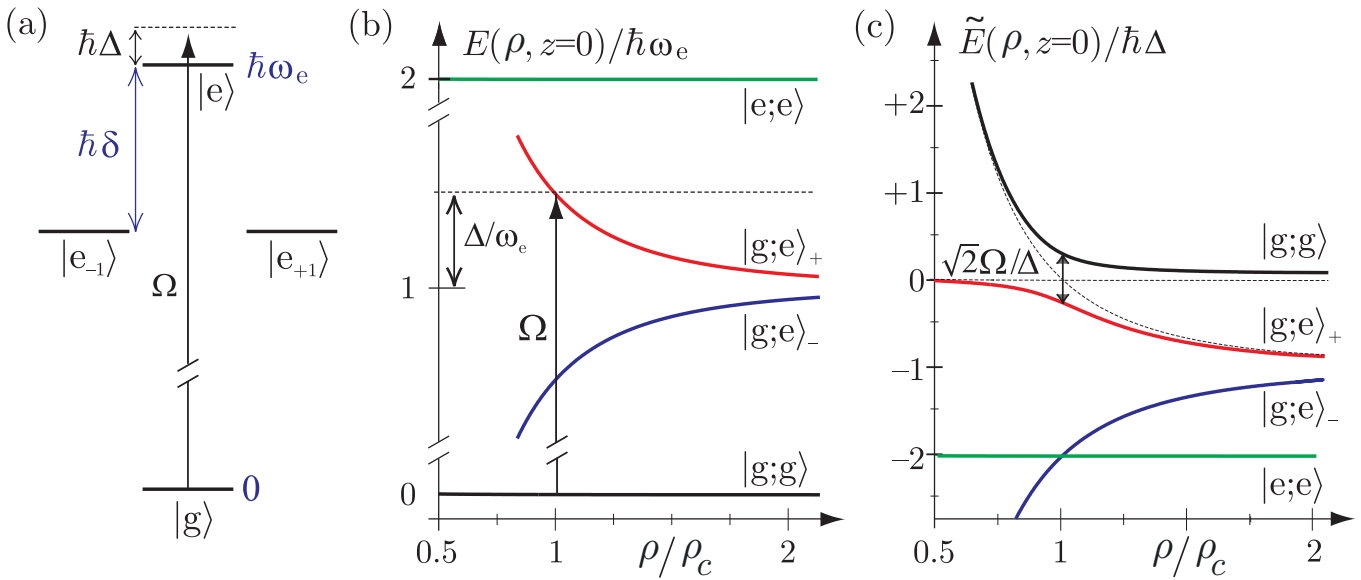}
\end{center}
\caption{Design of the step-like potential of Fig.~\ref{fig:fig1}(b): (a) Rotational spectrum of a molecule in a weak DC field. The DC field splits the $(J=1)$-manifold by an amount $\hbar \delta$.
The linearly-polarized microwave transition with detuning $\Delta$ and Rabi frequency
$\Omega$ is shown as an arrow. (b) BO-potentials for the internal
states for $\Omega=0$ (bare potentials), where
$|g;e\rangle_\pm\equiv(|g;e\rangle\pm|e;g\rangle)/\sqrt{2}$. The resonant Condon point $\rho_C$ is indicated by an arrow. (c) AC-field-dressed BO-potentials. The dressed grounstate potential has the largest energy.}\label{fig20000}
\end{figure*}

By combining DC and AC fields to dress the manifold of rotational excitations we can design effective interaction potentials $V^{{\rm 3D}}\left({\bf r}_{i}\!-\!{\bf
r}_{j}\right)$ with (essentially) any given shape as  a function of distance.
For example, the addition of a single linearly-polarized AC
field to the configuration of Fig.~\ref{fig:fig1}(a) leads to the realization of the
2D ``step-like'' potential of
Fig.~\ref{fig:fig1}(b), where the character of the repulsive
potential varies considerably in a small region of space.
The derivation of this effective 2D interaction is sketched in Fig.~\ref{fig20000} and it is discussed in more detail in Sect.~\ref{sec:secAC} below~\cite{Buechler07,Micheli07}. %, where it is assumed that a deep optical lattice confines the motion of the particles to the ($z=0$)-plane, as in Fig.~\ref{fig:fig1}(a).
The (weak) DC-field splits the first-excited rotational ($J=1$)-manifold of each molecule by an amount $\hbar \delta$, while a linearly polarized AC-field with Rabi frequency $\Omega$ is blue-detuned from the ($|g\rangle-|e\rangle$)-transition by $\hbar \Delta$, see Fig.~\ref{fig20000}(a). Because of $\hbar \delta$ and the choice of polarization, for distances $\rho \gg (d^2/\hbar \delta)^{1/3}$ the relevant single-particle states for the two-body interaction reduce to the states $|g\rangle$ and $|e\rangle$ of each molecule. Figure~\ref{fig20000}(b) shows that
the dipole-dipole interaction splits the excited state manifold of the two-body rotational spectrum, making the detuning $\Delta$ {\em position-dependent}. As a consequence, the combined energies of the bare groundstate of the two-particle spectrum and of a microwave photon become degenerate to the energy of a (symmetric) excited state at a characteristic resonant (Condon) point $\rho_C=(d^2/\hbar \Delta)^{1/3}$, which is represented by an arrow in Fig.~\ref{fig20000}(b). At this Condon point, an avoided crossing occurs in a {\em field-dressed} picture, and the new (dressed) groundstate potential inherits the character of the bare ground and excited potentials for distances $\rho\gg \rho_C$ and $\rho\ll \rho_C$, respectively. Consistently,
Fig.~\ref{fig20000}(c) shows that the dressed groundstate potential (which has the {\em largest energy}) is almost flat for $\rho\gg \rho_C$ and it is strongly repulsive as $1/\rho^3$ for $\rho\ll \rho_C$, which corresponds to the realization of the step-like potential of Fig.~\ref{fig:fig1}(b).
We remark that, due to the choice of polarization, this strong repulsion is present only in the plane $z=0$, while for $z\neq0$ the groundstate potential can turn into attractive. Thus, here the optical confinement along $z$ of Fig.~\ref{fig:fig1}(a) is necessary to ensure the stability of the collisional setup.

%It is
%shown in Sect.~\ref{sec:secAC} below that the change in character of the
%ground-state interaction potential corresponds to the absorption of
%a microwave photon at a specific Condon point $r_{\rm C}=(d^2/\hbar \Delta)^{1/3}$ in a {\em
%dressed} picture, with $\Delta$ the detuning of the microwave field.
%I FEEL THAT THE BEGINNING OF THIS SECTION MUST BE REWORKED AS THE ESSENTIAL PHYSICAL PICTURE IS NOT MADE EXPLICIT ENOUGH. WE MUST EXPLAIN THAT IN THE PRESENCE OF THE AC FIELD THE GROUND STATE PLUS A MICROWAVE PHOTON IS DEGENERATE WITH AN EXCITED STATE, INDUCING A DRESSED POTENTIAL. I WONDER IF WE NEED THE DIPOLE INTERACTION FIGURE HERE TO EXPLAIN THIS. THE NOTION OF A CONDON POINT IS NOT KNOW TO THE AUDIENCE. I AM NOT SURE ONE UNDERSTANDS THE SYMBOLS LIKE DELTA.

The interactions in the presence of a single AC
field are studied in quite detail in Ref.~\cite{Micheli07}, where
it is shown that in the absence of external confinement, this case
is analogous to the (3D) optical blue-shielding developed in the
context of ultracold collisions of neutral
atoms~\cite{Zilio96,Napolitano97,Weiner99}, however with the
advantage of the long lifetime of the excited rotational states of
the molecules, as opposed to the electronic states of cold atoms. The strong inelastic
losses observed in 3D collisions with cold
atoms~\cite{Zilio96,Napolitano97,Weiner99} can be avoided via a judicious choice of the
field's polarization, eventually combined with a tight confinement
to a 2D geometry (as e.g. for the case of Fig.~\ref{fig20000} above). For example, in Ref.~\cite{Gorshkov08} it is shown that for interactions in the presence of a DC field and of a circularly polarized AC field the {\em attractive} time-averaged interaction due to the rotating (AC-induced) dipole moments of the molecules  allows for the cancelation of the total dipole-dipole interaction. The residual interactions remaining after this cancelation are {\em purely repulsive 3D interactions} with a characteristic van-der-Waals behavior $V_{\rm eff}^{\rm 3D}({\bf r})\sim (d^4/\hbar \Delta)/r^6$. This 3D repulsion provides for a shielding of the inner part of the interaction potential and thus it will strongly suppress inelastic collisions in experiments. This will possibly lead to the realization of quantum degenerate systems of molecules and maybe of tightly-packed crystalline structures in {\em three-dimensions}~\cite{Gorshkov08}.

A cancelation of the leading effective two-body interaction similar to the one sketched above in a {\em dense} cloud of molecules can lead to the realization of systems where the effective {\em three-body} interaction $W^{\rm 3D}({\bf r}_i,{\bf r}_j,{\bf r}_k)$ of Eq.~\eqref{effint1} dominates over the two-body term $V^{\rm 3D}({\bf r}_i-{\bf r}_j)$ and determines the groundstate properties of the system.
This is interesting, since model Hamiltonians with strong
three-body and many-body interactions have attracted a lot of interest in the
search for microscopic Hamiltonians exhibiting exotic ground state
properties. Well known examples are the fractional quantum Hall
states described by the Pfaffian wave functions which appear as
ground states of a Hamiltonian with three-body interactions
\cite{moore91,fradkin98,cooper04}. These topological phases admit
anyonic excitations with non-abelian braiding statistic. Three-body
interactions are also an essential ingredient for systems with a low
energy degeneracy characterized by string
nets,\cite{levin05,fidkowski06} which play an important role in
models for non-abelian topological phases.
This possibility of realizing a Hamiltonian where the two-body interaction can be manipulated independently of the three-body term has been studied in Ref.~\cite{BuchlerNature07}. There, it is shown that a stable system where particles interact via purely repulsive three-body interactions can be realized by combining the setup above with the tight optical confinement provided by an optical lattice. In fact, the latter serves the two-fold purpose of ensuring the collisional stability of the setup and of defining a characteristic length scale (the lattice spacing) where the exact cancelation of the two-body term occurs. The details of this derivation are given in Sect.~\ref{sec:sec3Body} below, in connection with the derivation of an extended Hubbard model with three-body interactions introduced in the following Sect.~\ref{sec:secHubLatt}. %This collisionally stable setup leads to the realization of the following extended lattice models with exotic groundstate properties.

\subsection{Hubbard lattice models}\label{sec:secHubLatt}

%\begin{figure}[htb]
%\begin{center}
%\includegraphics[width=\columnwidth]{figs/fig03v1.eps}%[width=\columnwidth]{fig2-1.eps}
%\end{center}
%\caption{Strengths of the
%dominant three-body interactions $W_{i j k}$ appearing in the
%Hubbard model for different lattice geometries: (a)
%one-dimensional setup; (b) two-dimensional square lattice; (c)
%two-dimensional honeycomb lattice.  The characteristic energy scale
%$W_{0}= \gamma_{2} D R_{0}^6/a^6$ is discussed following
%Eq.~(\ref{threebodystrength}).} \label{fig2}
%\end{figure}

Hubbard Hamiltonians are model Hamiltonians describing the low-energy physics of interacting fermionic and bosonic particles in a lattice~\cite{Hubb}, and have the general tight-binding form
\begin{equation}\label{Hubbard10}
 H= - \sum_{i, j, \sigma } J_{i j}^{\sigma} b^{\dag}_{i,\sigma} b_{j,\sigma} + \sum_{i , j,\sigma,\sigma'}
\frac{U_{i j}^{\sigma \sigma'}}{2} n_{i,\sigma} n_{j,\sigma'}.
\end{equation}
Here $b_{i,\sigma}$ ($b^{\dag}_{i,\sigma}$) are destruction (creation) operators for a particle at site $i$ in the internal state $\sigma$, $J_{i j}^{\sigma}$ describes the coherent hopping of a particle from site $i$ to site $j$ (typically the nearest neighbor), and $U_{i j}^{\sigma \sigma'}$ describes the onsite ($i=j$) or offsite ($i\neq j$) two-body interactions between particles, with $n_{i,\sigma}= b^{\dag}_{i,\sigma} b_{i,\sigma}$. Hubbard models have a long history in condensed matter, where they have been used as tight-binding approximations of strongly correlated systems. For example, for particles being electrons in a crystal hopping from the orbital of a given atom to that of its nearest neighbor, $\sigma$ represents the electron spin. A (fermionic) Hubbard model comprising electrons in a 2D lattice with interspecies onsite interactions is thought to be responsible for the high-temperature superconductivity observed in cuprates~\cite{Cup1}.\\

\begin{figure*}
  \begin{center}
    \includegraphics[width=\textwidth]{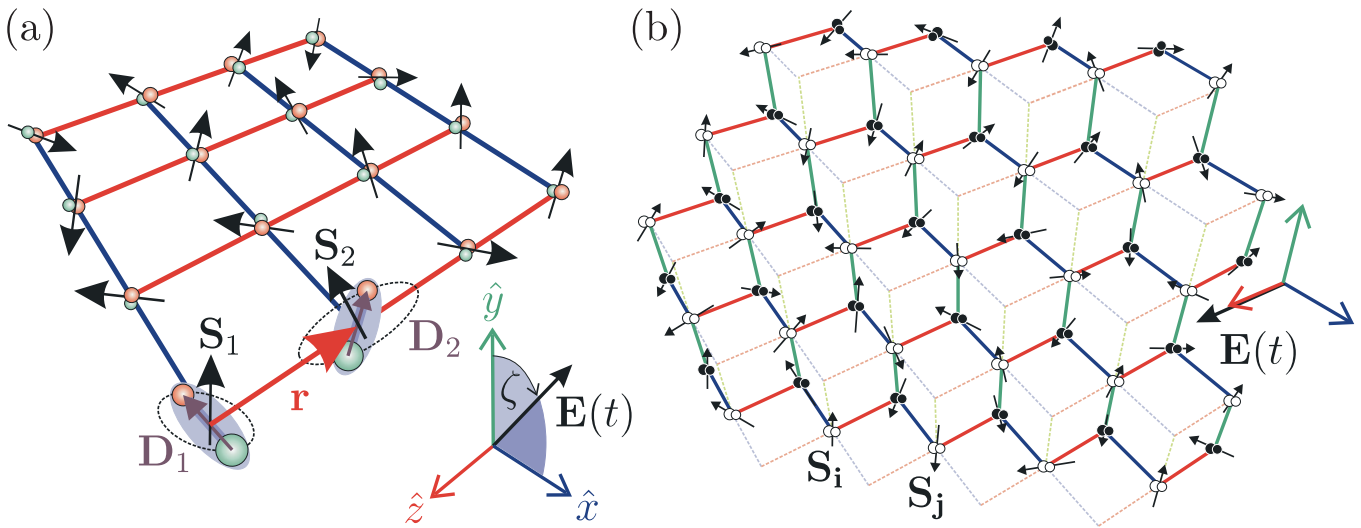}%{figs/Figure_4.eps}%{Rev23_Fig1.eps}
    \caption{\label{fig:1}Example anisotropic spin models
    that can be simulated with polar molecules trapped in optical lattices.
   (a) Square lattice in $2$D with nearest neighbor orientation dependent Ising interactions
   along $\hat{x}$ and $\hat{z}$.  Effective interactions between the spins
      ${\bf S}_1$ and ${\bf S}_2$ of the molecules in their rovibrational ground states
      are generated with a
      microwave field ${\bf E}(t)$ inducing
      dipole-dipole interactions between the molecules
      with dipole moments ${\bf D}_1$ and ${\bf D}_2$, respectively.
     (b) Two staggered triangular lattices with nearest neighbors oriented along
     orthogonal triads. The interactions depend on the orientation of the links with
     respect to the electric field.  (Dashed lines are included for perspective.)
    }
  \end{center}
\end{figure*}

In recent years, Hubbard models have been shown to properly describe the low-energy physics of interacting bosonic and fermionic atoms trapped at the bottom of an optical lattice~\cite{jaksch98,Hofs02}. The resulting low-energy Hamiltonians are of the form of Eq.~\eqref{Hubbard10} above. Spectacular experiments with ultracold atoms have lead to the realization of the superfluid/Mott-insulator quantum phase transition for bosonic atoms~\cite{Bloch02,Spielman06}, and great experimental progresses with fermions promise to solve the phase diagram of the fermionic Hubbard model in 2D by performing an {\em analog quantum simulation} of Eq.~\eqref{Hubbard10} with two-species cold fermions~\cite{Ess1,Ess2}.

Since the interactions between cold atoms are short-ranged, in these systems Hubbard Hamiltonian typically have {\em onsite} interactions only [$U_{i,i}^{\sigma,\sigma'}$ in Eq.~\eqref{Hubbard10}]. However, it has been shown that the presence of moderately long-range interactions in Eq.~\eqref{Hubbard10}, such as nearest-neighbor interactions, can lead to interesting phases such as checkerboard solids and 2D supersolids~\cite{Opt6,Opt7}. Polar molecules in optical lattices can provide for offsite interactions~\cite{Opt1,Opt2,Opt3,Opt4,Opt5} which are strong, of the order of hundreds of kHz, and long-range, i.e. they decay with distance as $1/|i-j|^3$. Due to these strong interactions, two molecules cannot hop onto the same site, and thus the particles are treated as effectively "hard-core".
\\

An intriguing possibility is offered by the interaction engineering discussed above in the context of the realization of effective lattice models where particles interact via exotic (extended) Hubbard Hamiltonians. An example of this is given in Ref.~\cite{BuchlerNature07}, where it is shown how to engineer the following Hubbard-like Hamiltonian
\begin{equation}\label{Hubbard}
 H= - J \sum_{\langle i j \rangle} b^{\dag}_{i} b_{j} + \sum_{i \neq j}
\frac{U_{i j}}{2} n_{i} n_{j} + \sum_{i \neq j \neq k} \frac{W_{i j
k}}{6} n_{i} n_{j} n_{k},
\end{equation}
where ${W_{i j k}}n_{i} n_{j} n_{k}$ is an offsite three-body term. The latter is tunable independently of the two-body term $U_{i j} n_{i} n_{j}$, to the extent that it can be made to dominate the dynamics and determine the groundstate properties of the system.
In contrast to the common approach to derive effective many-body  terms from
Hubbard models involving two-body interactions, which are obtained
in a $J\ll U$ perturbation theory, and are thus necessarily
small~\cite{tewari06}, the derivation of the Hubbard model Eq.~(\ref{Hubbard}) is based directly on the effective many-particle potential Eq.~(\ref{effint1}).
Thus, all the energy scales in Eq.~\eqref{Hubbard} can be tuned independently, which allows to obtain comparatively large hopping rates determining the time and temperature scales to observe exotic quantum phases. This is important, since e.g. in 1D analytical calculations suggest that the Hamiltonian Eq.~(\ref{Hubbard}) has a rich groundstate phase-diagram, supporting valence-bond, charge-density-wave and superfluid phases~\cite{BuchlerNature07}. In Sect.~\ref{sec:secApplications} below we provide the microscopic derivation of the effective interaction potentials of Eq.~\eqref{Hubbard}.

\subsection{Lattice Spin models}\label{sec:secLattSpin}

The Hamiltonian Eq.~\eqref{eq:eqGeneral} can be generalized to include other internal degrees of freedom for each molecule in addition to rotation. This offers new possibilities to engineer effective interactions and novel many-body phases. For example, the addition of a spin-1/2 (qubit) degree of freedom to polar molecules trapped into an optical lattice allows to construct a {\em complete toolbox} for the simulation of any permutation symmetric lattice spin models~\cite{micheli06}.
Lattice spin models are ubiquitous in condensed matter physics where
they are used as simplified models to describe the characteristic
behavior of more complicated interacting physical systems.

The basic building block is a system of two polar molecules strongly
trapped at given sites of an optical lattice, where the spin-$1/2$
(or qubit) is represented by a single electron outside a closed
shell of a $^{2} \Sigma_{1/2}$ heteronuclear molecule in its rotational ground state, as provided e.g. by alkaline-earth monohalogenides. As dicussed above, heteronuclear molecules have large permanent electric dipole moments, which are responsible for strong, long-range and anisotropic dipole-dipole interactions, whose \emph{spatial dependence} can be manipulated using microwave fields. Accounting for the spin-rotation splitting of
molecular rotational levels these dipole-dipole
interactions can be made \emph{spin-dependent}. General lattice spin models are then
readily built from these binary interactions. Although in this review we will present results for spin-1/2 models only, we notice that the inclusion of hyperfine effects offers extensions to spin systems with larger spin. For example, the design of a large class of spin-1 interactions for polar molecules has been shown in Ref.~\cite{brennenNJP07}, which allows e.g. for the realization of a generalized Haldane model in 1D~\cite{Haldane}.

Two highly anisotropic models with spin-$1/2$ particles that can be simulated are illustrated in Figs.~\ref{fig:1}(a) and \ref{fig:1}(b) respectively.  The first takes place on a square $2$D lattice with nearest
  neighbor interactions
\begin{equation}
  H_{\rm spin}^{({\rm I})}= \sum_{i=1}^{\ell -1}\sum_{j=1}^{\ell -1} J (\sigma^z_{i,j}\sigma^z_{i,j+1}+\cos\zeta\sigma^x_{i,j}\sigma^x_{i+1,j}).
  \label{Ioffe}
\end{equation}
Introduced by Dou\c{c}ot {\it et al.} \cite{Duocot:05} in the
context of Josephson junction arrays, this model (for $\zeta\neq
\pm\pi/2$) admits a 2- fold degenerate ground subspace that is
immune to local noise up to $\ell$-th order and hence is a good
candidate for storing a protected qubit.

The second, occurs on a bipartite lattice constructed with two $2$D
triangular lattices, one shifted and stacked on top of the other.
The interactions are indicated by nearest neighbor links along the
$\hat{x}, \hat{y}$ and $\hat{z}$ directions in real space:
\begin{equation}
H_{\rm spin}^{({\rm II})}=J_{\perp}\sum_{x-{\rm
links}}\sigma^x_j\sigma^x_k+ J_{\perp} \sum_{y-{\rm
links}}\sigma^y_j\sigma^y_k+ J_z\sum_{z-{\rm
links}}\sigma^z_j\sigma^z_k. \label{Kit}
\end{equation}
This model has the same spin dependence and nearest neighbor graph
as the model on a honeycomb lattice introduced by Kitaev
\cite{Kitaev:05}. He has shown that by tuning the ratio of
interaction strengths $|J_{\perp}|/|J_z|$ one can tune the system
from a gapped phase carrying abelian anyonic excitations to a
gapless phase which in the presence of a magnetic field becomes
gapped with non-abelian excitations. In the regime
$|J_{\perp}|/|J_z|\ll 1$ the Hamilonian can be mapped to a model
with four body operators on a square lattice with ground states that
encode topologically protected quantum memory \cite{DKL:03}. One
proposal \cite{Duan:03} describes how to use trapped atoms in spin
dependent optical lattices to simulate the spin model $H_{\rm
spin}^{({\rm II})}$. There the induced spin couplings are obtained
via spin dependent collisions in second order tunneling processes.
Larger coupling strengths as provided by polar molecules are desirable.
In both spin models (${\rm
I}$ and ${\rm II}$) above, the signs of the interactions are
irrelevant although one is able to tune the signs if needed.

\subsection{Hubbard models in self-assembled dipolar lattices}\label{sec:secFloat}

\begin{figure}[htb]
\begin{center}
\includegraphics[width=0.9\columnwidth]{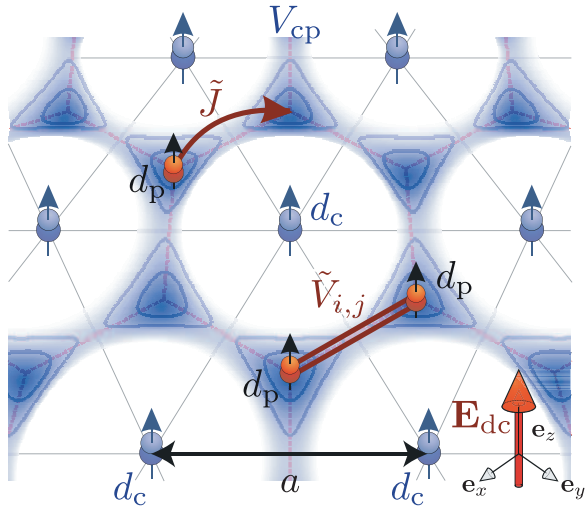}
\end{center}
\caption{Floating lattices of dipoles: A self-assembled crystal of polar molecules with dipole moment $d_{\rm c}$ provides a 2D periodic honeycomb lattice $V_{{\rm cp}}$ (darker shading
corresponds to deeper potentials) for
extra molecules with dipole $d_{\rm p} \ll d_{\rm c}$ giving rise to a lattice model with hopping
$\tilde{J}$ and long-range interactions $\tilde{V}_{i,j}$. } \label{fig2}
\end{figure}

In Hubbard models with cold atoms or molecules in optical lattices there is no phonon degrees of freedom corresponding to an intrinsic dynamics of the lattice, as the back action on the optical potentials is typically negligible. Thus atomic and molecular Hubbard models allow the study of strong correlations in the absence of phonon effects. However, the simulation of models where the presence of (crystal) phonons strongly affects the (Hubbard) dynamics of the particles remains a challenge. These models are of fundamental interest in condensed matter physics, where they describe polaronic and/or superconducting materials~\cite{Alexandrov}.  In the context of atoms, one example is to immerse atoms moving on a lattice into a BEC of a second atomic species, representing a bath of Bogoliubov excitations. A second example is a self-assembled floating lattice of molecules  as discussed in  Sect.~\ref{sec:secSelfAssembled}, which provides a periodic potential for extra atoms or molecules, whose dynamics can be again described in terms of a Hubbard model \cite{Pupillo08}. Phonon degrees of freedom enter as vibrations of the dipolar lattice.

The Hamiltonian for extra atoms or molecules in a self-assembled dipolar lattice is
\begin{eqnarray}
H & = & -J\sum_{<i,j>}c_{i}^{\dagger}c_{j}
+\tfrac{1}{2}\sum_{i,j}V_{ij}c_{i}^{\dagger}c_{j}^{\dagger}c_{j}c_{i}\nonumber \\
 & + & \sum_q \hbar \omega_q a_{q}^{\dagger}a_{q}+\sum_{q,j}M_{ q}e^{i{\bf q}\cdot{\bf R}_{j}^{0}}c_{j}^{\dagger}c_{j}(a_{q}
 +a_{-q}^{\dagger}).\nonumber \label{eq:eqSmallPolaron}\end{eqnarray}
Here, the first and second terms define a Hubbard-like Hamiltonian for the extra-particles of the form of Eq.~\eqref{Hubbard10}, where the operators $c_{i}$ ($c_{i}^{\dagger}$) are
destruction (creation) operators of the extra-particles. However, the third and fourth terms describe the acoustic phonons of the crystal and the coupling of the extra-particles to the crystal phonons, respectively. Here, $a_q$ destroys a phonon  of quansimomentum ${\bf q}$ in the mode $\lambda$. Tracing over the phonon degrees of freedom in a strong coupling limit provides effective Hubbard models for the extra-particles dressed by the crystal phonons
\begin{equation}
\tilde{H}=-\tilde{J}\sum_{
<i,j>}c_{i}^{\dagger}c_{j}+\tfrac{1}{2}
\sum_{i,j}\tilde{V}_{ij}c_{i}^{\dagger}c_{j}^{\dagger}c_{j}c_{i}\nonumber
\end{equation}
The hopping of a dressed extra-particle between the minima of the periodic potential occurs at a rate $\tilde{J}$, which is exponentially suppressed due to the co-propagation of the lattice distortion, while offsite particle-particle interactions $\tilde{V}_{i,j}$ are now a combination of direct particle-particle interactions and interactions mediated by the coupling to phonons. The setup we have in mind is depicted in Fig.~\ref{fig:fig2}(b), where extra-particles which are molecules with a dipole moment $d_{\rm p}\ll d_{\rm c}$ interact repulsively with the crystal molecules, and thus see a periodic (honeycomb) lattice potential.

The distinguishing features of this realization of
lattice models are: (i) Dipolar molecular crystals constitute an
array of microtraps with its own quantum dynamics represented by
phonons (lattice vibrations), while the lattice spacings are tunable
with external control fields, ranging from a $\mu$m down to the
hundred nm regime, i.e. potentially smaller than for optical
lattices. (ii) The motion of the extra particles is governed by an
interplay of \emph{Hubbard (correlation) dynamics} in the lattice
and \emph{coupling to phonons}. The tunability of the lattice allows
to access a wide range of Hubbard parameters and phonon couplings.
Compared with optical lattices, for example, a small scale lattice
yields significantly enhanced hopping amplitudes, which set the
relevant energy scale for the Hubbard model, and thus
also the temperature requirements for realizing strongly correlated
quantum phases.

\section{Engineering of interaction potentials}\label{sec:secEngine}

In this section we show in some detail how to realize a collisionally stable 2D setup where particles interact via a purely repulsive $1/r^3$ potential, by using a combination of a DC field and of tight optical confinement in the field's direction. We then sketch how to design more complicated interactions using a combination of AC, DC, and optical fields, by focussing on the step-like potential of Fig.~\ref{fig:fig1}(b). This engineering of interaction potentials is at the core of the realization of the strongly correlated phases and quantum simulations discussed in Sect.~\ref{sec:secApplications} below.

\subsection{Molecular Hamiltonian}\label{sec:secMolecularHam}

We consider spin-less polar molecules in their electronic and
vibrational ground-state, with spectral notation $X ^1\Sigma(0)$. In
the following, we are interested in manipulating their rotational
states using DC and AC electric fields and in confining their motion
using a (optical) far-off-resonance trap (FORT). The application of
these external fields will serve as a key element to engineer
effective interaction potentials between the molecules.

The low energy effective Hamiltonian for the external motion and
internal rotational excitations of a single molecule is
\begin{eqnarray}
H(t) = \frac{{\bf p}^2}{2m} + H_{\rm rot} + H_{\rm DC} + H_{\rm
AC}(t) + H_{\rm opt}({\bf r}), \label{eq:basic}\end{eqnarray} where
${\bf p}^2/2m$ is the kinetic energy for the center-of-mass motion
of a molecule of mass $m$, $H_{\rm rot}$ accounts for the rotational
degrees of freedom, while the  terms $H_{\rm DC}$, $H_{\rm AC}(t)$
and $H_{\rm opt}({\bf r})$ refer to the interaction with electric DC
and AC (microwave) fields and to the optical trapping of the
molecule in the ground electronic-vibrational manifold,
respectively. In the following we consider {\em tight} harmonic
optical traps with a frequency $\omega_{\perp}= 2\pi\times 150$kHz, which is the same for all the relevant rotational states of the molecule. That is, we neglect possible tensor-shifts induced by the optical trapping in the energies of the excited rotational states of the molecules, which in general can be compensated for by an appropriate choice of additional laser fields \cite{Micheli07}. Thus, for a confinement along ${\bf e}_z$, $H_{\rm opt}({\bf r})$ reads $H_{\rm opt}({\bf r})=m
\omega_{\perp}^2 z^2/2$, independent of the internal (rotational)~\cite{all7,Friedrich95}.
\\

{\em Rotational spectrum} :- The term $H_{\rm rot}$ in
Eq.~\eqref{eq:basic} is the Hamiltonian for a rigid spherical
rotor~\cite{Herzberg50}
\begin{eqnarray}
  H_{\rm rot}=B{\bf J}^2,\label{eq:rotor}
\end{eqnarray}
which accounts for the rotation of the internuclear axis of a
molecule with total internal angular momentum ${\bf
J}$~\cite{Herzberg50,Brown03,Judd75}. Rotations are the
lowest-energy internal excitations of the molecule. Here $B$ is the
rotational constant for the electronic-vibrational ground state,
which is of the order of $B\sim h~10~{\rm GHz}$~\cite{NISTDataWeb}.
We denote the energy eigenstates of Eq.~\eqref{eq:rotor} by
$\ket{J,M}$, where $J$ is the quantum number associated with the
total internal angular momentum and $M$ is the quantum number
associated with its projection onto a {\em space-fixed} quantization
axis. The excitation spectrum is $E_{J}=B J(J+1)$, which is
anharmonic. Each $J$-level is $(2J+1)$-fold
degenerate.\\

A polar molecule has an electric dipole moment, ${\bf d}$, which
couples its internal rotational levels and for $\Sigma$-molecules is
directed along the internuclear axis ${\bf e}_{ab}$, i.e. ${\bf
d}=d{\bf e}_{ab}$. Here, $d$ is the ``permanent'' dipole moment of a
molecule in its electronic-vibrational ground-state. This dipole
moment is responsible for the dipole-dipole interaction between two
molecules.

The spherical component $d_q={\bf e}_q\cdot{\bf d}$ of
the dipole operator on the {\em space-fixed} spherical basis $\{{\bf
e}_{-1},{\bf e}_0,{\bf e}_1\}$, with ${\bf
  e}_{q=0}\equiv{\bf e}_z$ and ${\bf e}_{\pm 1}=\mp({\bf e}_x\pm i{\bf
  e}_y)/\sqrt{2}$
couples the rotational states $\ket{J,M}$ and
$\ket{J\pm1,M+q}$ according to
\begin{eqnarray}
  \bra{J\pm1,M+q}d_q\ket{J,M} =
  d(J,M;1,q|J\pm1,M+q)\times\nonumber\\
   \times (J,0;1,0|J\pm1,0)\sqrt{\frac{2J+1}{2(J\pm1)+1}},\nonumber
\end{eqnarray}
where $(J_1,M_1;J_2,M_2|J,M)$ are the Clebsch-Gordan-coefficients. This means that for a spherically-symmetric system the eigenstates of the rotor have
no net dipole-moment, $\bra{J,M}{\bf d}\ket{J,M}=0$.
However, the dipole coupling to an external electric field breaks this spherical
symmetry by aligning each molecule along the field's direction. This
induces a dressing of the rotational energy levels of the molecule,
and a corresponding finite dipole moment in each rotational state,
as explained below.\\

{\em Coupling to external electric fields} :- The terms $H_{\rm DC}$
and $H_{\rm AC}(t)$ in Eq.\eqref{eq:basic} are the electric dipole
interaction of a molecule with an external DC electric field ${\bf
E}_{\rm DC}=E_{\rm DC} {\bf e}_z$ directed along ${\bf
e}_0\equiv{\bf e}_z$, and with AC microwave fields ${\bf E}_{\rm
  AC}(t)=E_{\rm AC}e^{-i\omega t}{\bf e}_q+{\rm c.c.}$, which are linearly ($q=0$)
or circularly polarized ($q=\pm1$) relative to ${\bf e}_z$,
respectively. Here we have neglected the spatial
dependence of $E_{\rm AC}$ since in the following we are interested in
dressing the rotational states of the molecules with microwave
fields, whose wavelengths are of the order of centimeters, and thus much larger than
the size of our system. Then, the terms $H_{\rm DC}$
and $H_{\rm AC}(t)$ in Eq.~\eqref{eq:basic} read
\begin{subequations}
  \begin{eqnarray}
    H_{\rm DC} &=& -{\bf
      d}\cdot{\bf E}_{\rm DC} = - d_0 E_{\rm DC},\label{eq:DCelectric}\\
    H_{\rm AC}(t) &=& -{\bf
      d}\cdot{\bf E}_{\rm AC}(t) = - d_q E_{\rm AC}e^{-i\omega
      t}+{\rm h.c.}\label{eq:acelectric}.\nonumber\\
  \end{eqnarray}
\end{subequations}
\\

\begin{figure}[htbp]
  \begin{center}
    \includegraphics[width=\columnwidth]{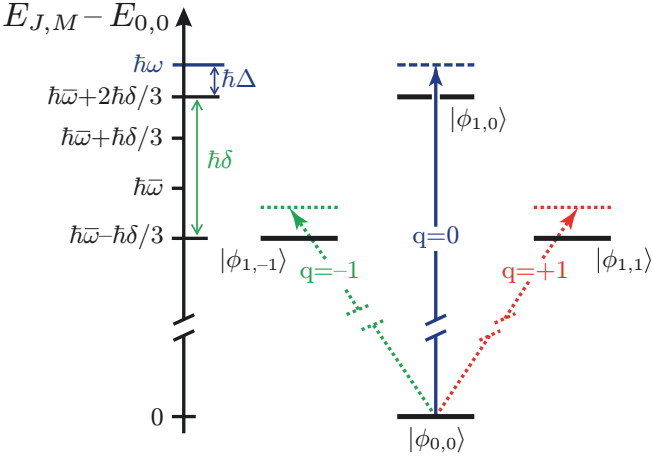}%{figs/fig04.eps}%{figs/Figure_5.eps}
  \end{center}
  \caption{\label{fig:fig2}Solid lines: Energies $E_{J,M}$ (left)
    and states $\ket{\phi_{J,M}}$ (right) with $J=0,1$,
    for a molecule in a weak DC electric field ${\bf E}_{\rm DC}=E_{\rm DC}{\bf e}_0$
     with $\beta\equiv d E_{\rm DC}/B \ll
      1$. The DC-field-induced splitting $\hbar \delta$
      and the average energy separation
      $\hbar \bar{\omega}$ are $\hbar \delta=3B\beta^2/20$
      and $\hbar \bar{\omega}=2B+B\beta^2/6$, respectively. Dashed
      and dotted lines: Energy levels for a molecule in
      combined DC and AC fields (The AC-Stark shifts of the dressed states are not shown).
      Dashed line: The AC field is monochromatic, with
      frequency $\omega$, linear polarization $q=0$, and detuning $\Delta=\omega-(\bar{\omega}
      +2 \delta/3)>0$. Dotted lines: Schematics of energy levels
      for an AC-field with polarization $q=\pm1$ and frequency $\omega'\neq\omega$.}
\end{figure}

In the presence of a single DC electric field ${\bf E}_{{\rm DC}}$
(${\bf E}_{{\rm AC}},\omega_{\perp}$=0), the internal Hamiltonian is
that of a rigid spherical pendulum \cite{Herzberg50}
%\begin{eqnarray}
 $ H = H_{\rm rot}+H_{\rm DC} = B{\bf J}^2 - d_0 E_{\rm
    DC}, \label{eq:pendulum}$
%\end{eqnarray}
which conserves the projection of the angular momentum $J$ on the
quantization axis, i.e. $M$ is a good quantum number. Thus, the
energy eigenvalues and eigenstates are labeled as $E_{J,M}$ and
$\ket{\phi_{J,M}}$, respectively, where each eigenstate
$\ket{\phi_{J,M}}$ is a superposition of various states $\ket{J,M}$
mixed by the electric dipole interaction.

The effects of a DC electric field on a single polar molecule are
shown in figure Fig.~\ref{fig:fig2}, and they amount to: (a) split the
$(2J+1)$-fold degeneracy in the rotor spectrum, and (b) align the
molecule along the direction of the field. The latter corresponds to
inducing a finite dipole moment in each rotational state. For weak
fields $\beta\equiv d E_{\rm DC} /B \ll1$, the state $\ket{\phi_{J,M}}$ and
its associated induced dipole moment approximately read
\begin{eqnarray}\label{eq:eqEigenEDC}
  \ket{\phi_{J,M}} &=& \ket{J,M} -
  \frac{\beta}{2}\frac{\sqrt{J^2-M^2}}{\sqrt{J^3(2J+1)}}
\ket{J-1,M}+\nonumber\\
&&+
\frac{\beta}{2}\frac{\sqrt{(J+1)^2-M^2}}{\sqrt{(J+1)^3(2J+1)}}\ket{J+1,M},
\end{eqnarray}
and
\begin{eqnarray}
\bra{\phi_{J,M}}{\bf d}\ket{\phi_{J,M}} =  d\beta
  \frac{3M^2/J(J+1)-1}{(2J-1)(2J+3)}{\bf e}_0,\nonumber
\end{eqnarray}
respectively. Thus, the ground state
acquires a finite dipole moment $
  \bra{\phi_{0,0}}d_0\ket{\phi_{0,0}}=d\beta/3$ along the field
axis, which is at the origin of ground-state dipole-dipole
interactions between polar molecules.

We notice that the for a typical rotational constant, $B\sim
h~10~{\rm GHz}$, and a dipole-moment $d\sim 9~{\rm Debye}$ the
condition $\beta \ll 1$ corresponds to considering DC fields (much)
weaker than $B/d\sim 2~{\rm kV/cm}$.\\

Individual transitions of the internal Hamiltonian can be addressed
by applying one (or several non-interfering) microwave field ${\bf
E}_{\rm AC}(t)$, which can be {\em linearly} or {\em circularly}
polarized. This is shown in Fig.~\ref{fig:fig2} for transitions
coupling the $J=0$ and $J=1$ manifolds, where the Rabi frequency
$\Omega$ and the detuning $\Delta$ are $\Omega\equiv E_{\rm
AC}\bra{\phi_{1,q}}d_q\ket{\phi_{0,0}}/\hbar$ and
$\Delta\equiv\omega-(E_{1,q}-E_{0,0})/\hbar$, respectively.%

Dressed energy levels of a molecule are obtained by diagonalizing
the Hamiltonian $H=H_{\rm rot}+ H_{\rm DC} + H_{\rm AC}(t)$ in a
Floquet picture. That is, first, the Hamiltonian is expanded on the
basis $\ket{\phi_{J,M}}$, which diagonalizes the time-independent
part of $H$ as $H_{\rm rot}+H_{\rm
DC}=\sum_{J,M}\ket{\phi_{J,M}}E_{J,M}\bra{\phi_{J,M}}$, and then the
time-dependent wave-function is expanded in a Fourier series in the
AC frequency $\omega$. After applying a rotating wave approximation,
i.e. keeping only the energy conserving terms, one obtains a {\em
time-independent} Hamiltonian $\tilde H$, whose eigenvalues
correspond to the dressed energy levels \cite{Micheli07}.%

\subsection{Two molecules}\label{Sec:TwoMolecules}

We now consider the interactions of two polar molecules $j=1,2$ confined
to the $x-y$ plane by a tight harmonic trapping potential of
frequency $\omega_\perp$, directed along $z$. The interaction of the
two molecules at a distance ${\bf r}\equiv{\bf r}_2-{\bf r}_1=r{\bf
e}_{r}$ is described by the Hamiltonian
\begin{eqnarray}
  H(t)  =   \sum_{j=1}^{2} H_j(t) + V_{\rm dd}({\bf r}),\label{eq:eqTwoMol}
  \label{eq:eq10}
\end{eqnarray}
where $H_j(t)$ is the single-molecule Hamiltonian
Eq.~\eqref{eq:basic}, and $V_{\rm dd}({\bf r})$ is the dipole-dipole
interaction of Eq.~\eqref{eq:eqDipDip1}. %WE HAD THIS VDD BEFORE IN SEC II AND I WONDER IF WE CAN JUST REFER TO THIS EQUATION
%\begin{eqnarray}
%V_{\rm dd}({\bf r}) =  \frac{{\bf d}_1\cdot{\bf d}_2-3\left({\bf
%d}_1\cdot{\bf
%      e}_r\right)\left({\bf e}_r\cdot{\bf d}_{2}\right)}{r^3}.\label{eq:eqDipDip}
%\end{eqnarray}
%Here, ${\bf d}_j$ is the dipole operator of the molecule $j$, and
%${\bf e}_r\cdot{\bf d}_j$ is its projection onto the collision axis
%${\bf e}_r$.
%The terms $d_{q;j} \equiv{\bf e}_q\cdot{\bf d}_j$ are the spherical
%components of the projection of the dipole operator of molecule $j$
%onto the space-fixed frame ${\bf e}_q$.

In the absence of external fields $E_{\rm DC}=E_{\rm AC}=0$, the
interaction of the two molecules in their rotational ground state is
determined by the van-der-Waals attraction $V_{\rm vdW}\sim
C_{6}/r^{6}$ with $C_{6}\approx-d^{4}/6B$. This expression for
the interaction potential is valid outside of the molecular core
region $r>r_{B }\equiv (d^{2}/B)^{1/3}$, where $r_{B}$ defines the
characteristic length where the dipole-dipole interaction becomes
comparable to the splittings of the rotational levels. In the
following we show that it is possible to \emph{induce} and
\emph{design} interaction potentials which are long-range, by
dressing the interactions with appropriately chosen static and/or
microwave fields. In fact, the combination of the latter with
low-dimensional trapping allows to engineer effective potentials
whose {\em strength} and {\em shape} can be both tuned. The
derivation of the effective interactions proceeds in two steps: (i)
We derive a set of Born-Oppenheimer (BO) potentials by first
separating Eq.~\eqref{eq:eq10} into center-of-mass and relative
coordinates, and diagonalizing the Hamiltonian $H_{\rm rel}$ for the
relative motion for fixed molecular positions. Within an adiabatic
approximation, the corresponding eigenvalues
play the role of an effective %$V_{\mathrm{eff}}^{\rm3D}(\mathbf{r})$
3D interaction potential in a given state manifold dressed by the
external field. (ii) We eliminate the motional degrees of freedom in
the tightly confined $z$ direction to obtain an effective 2D
dynamics with interaction $V_{\rm eff}^{\rm 2D}(\boldrho)$.\\

In the following we show how to design interaction potentials,
presenting in some details the simplest case of a purely repulsive
$1/r^3$ potential in 2D, obtained using a static electric field
(Sect.~\ref{sec:secDC}). We then sketch how to design more elaborate
potentials using a combination of static and microwave fields
coupling the lowest rotor states of each molecule (Sect.~\ref{sec:secAC}).

\subsubsection{{\em {\bf Designing the repulsive $1/r^3$ potential in 2D}}}\label{sec:secDC}

\begin{figure*}[htbp]
  \begin{center}
    \includegraphics[width=\textwidth]{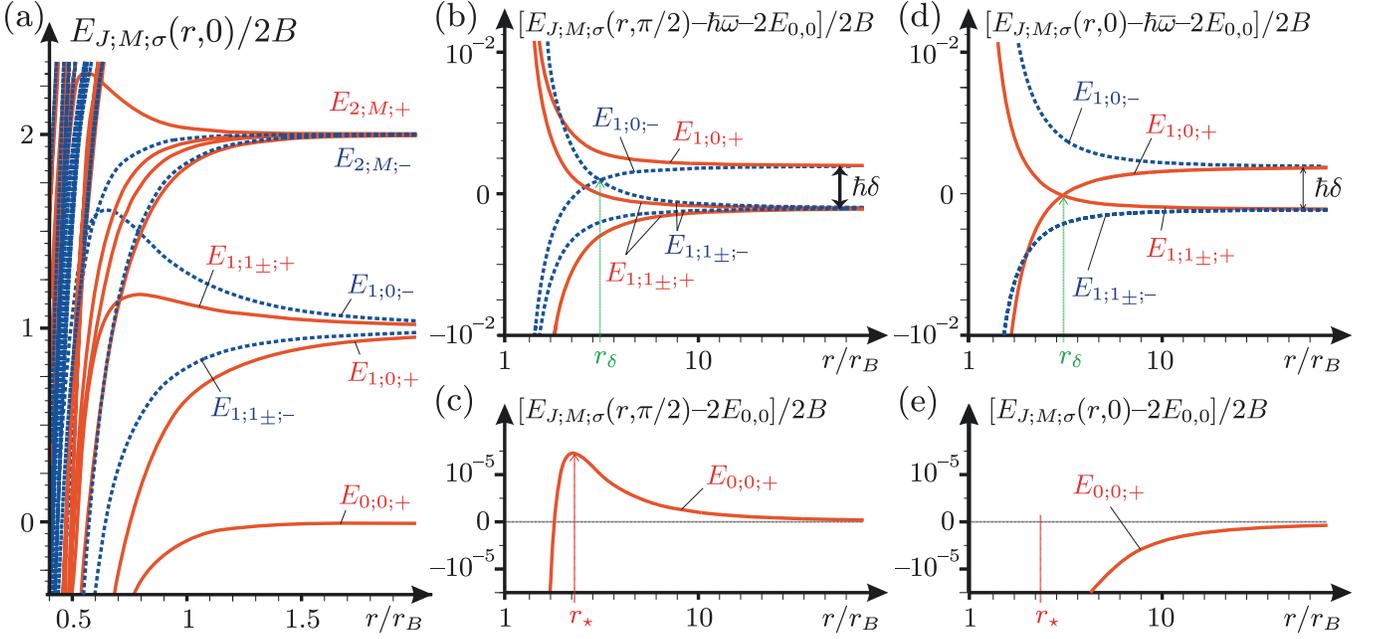}%{figs/fig06.eps}%{figs/Figure_6.eps}
    \caption{\label{fig:fig4}BO-potentials $E_{J;M;\sigma}(r,\vartheta)$ for two molecules colliding
      in the presence of a DC field, with
      $\beta\equiv d E_{\rm DC}/B=1/5$. The solid and dashed curves
      correspond to symmetric ($\sigma=+$) and antisymmetric ($\sigma=-$) eigenstates,
      respectively. (a): BO-potentials for the 16 lowest-energy eigenstates
      $E_{J;M;\sigma}(r,\vartheta)$. The molecular-core region is
      identified as the region $r<r_B=(d^2/B)^{1/3}$, while for $r\gg r_B $ the eigestates group into manifolds separated by one
      quantum of rotational excitation $2B$. (b) and (d): Blow-ups
      of the first-excited energy manifold of panel (a) in the region $r \gtrsim
      r_B$ for $\vartheta=\pi/2$ and $\vartheta=0$, respectively. Note the
      electric-field-induced splitting $\hbar \delta\equiv 3 B \beta^2 /20$.
      The distance $r_\delta$ where the dipole-dipole interaction
      becomes comparable to $\hbar \delta$ is $r_{\delta}=(d^2/\hbar\delta)^{1/3}$. (c) and (e): Blow-ups
      of the ground-state potential $E_{0,0;+}(r,\vartheta)$ of panel (a) in the region $r \gtrsim
      r_B$ for $\vartheta=\pi/2$ and $\vartheta=0$, respectively. The distance
      $r_\star$ of Eq.~\eqref{eq:rstar},
      where the dipole-dipole interaction becomes comparable to the van-der-Waals attraction
      is indicated. Note the repulsive (attractive) character of the potential for $\vartheta=\pi/2$
      ($\vartheta=0$) and $r>r_\star$.}
  \end{center}
\end{figure*}

{\em Collisions in a DC field}:- We consider a weak static electric field applied in the
$z$-direction ${\bf E}=E_{\rm DC}{\bf e}_0$ with $\beta=d E_{\rm DC}/B \ll
1$, and in the absence of optical trapping ($\omega_\perp=0$). The
effective interaction potentials for the collision of the two
particles can be obtained in the adiabatic approximation by
neglecting the kinetic energy and by diagonalizing the following
Hamiltonian $H_{\rm rel}$ for fixed particle positions
\begin{eqnarray}
H_{\rm rel}&=& \sum_{j=1}^{2} \left[ B{\bf J}_j^2 - E_{\rm DC}
d_{0;j}\right] + V_{\rm dd}({\bf r})\nonumber\\ &=& \sum_n
\ket{\Phi_n({\bf r})}E_n({\bf r})\bra{\Phi_n({\bf r})},
\end{eqnarray}
where $E_n({\bf r})$ and $\ket{\Phi_n({\bf r})}$ are the $n$th
adiabatic energy eigenvalues and two-particle eigenfunctions,
respectively. In the limit $r\rightarrow \infty$ the latter are
symmetrized products of the single-particle states
$\ket{\phi_{J_j,M_j}}_j$ of Eq.~\eqref{eq:eqEigenEDC}, % of Eq.~\eqref{eq:EigStatic},
while for
finite $r$ they are superpositions of several single-particle
states, which are mixed by the dipole-dipole interaction $V_{\rm
dd}({\bf r})$. The quantity $n\equiv(J;M;\sigma)$ is the collective
quantum number labeling the eigenvalues $E_n({\bf r})$, with
$J=J_1+J_2$ the total number of rotational excitations shared by the
two molecules, $M\equiv|M_1|+|M_2|$ the total projection of angular
momentum onto the electric field direction, and $\sigma=\pm$ the
permutation symmetry associated with the exchange of the two
particles. We note that, because of the presence of the DC field,
here $J$ is a simple label for the various energy manifolds, and not
a quantum number.

Since we are mainly interested in ground-state collisions, in the
following we restrict our discussion to the $J_j=0$ and 1 manifolds
of each molecule, which amounts to taking into account 16 rotational
two-particle states. Figure~\ref{fig:fig4} shows the corresponding
eigenvalues $E_n({\bf r})$ as a function of the interparticle
distance $r$, for $\beta=1/5$. The vector ${\bf r}$ is expressed in
spherical coordinates ${\bf r}=(r,\vartheta,\varphi)$, with $\vartheta$
and $\varphi$ the polar and azimuthal angles, respectively, and $z=r
\cos\vartheta$. Figure~\ref{fig:fig4}(a) shows that the energy spectrum has a markedly different behavior in the molecular core region $r<r_B$ and for $r>r_B$, with $r_B\equiv (d^2/B)^{1/3}$. In fact, for $r<r_B$ the energy spectrum is characterized by a series of level crossings and anti-crossings, which make the satisfaction of the adiabatic approximation generally impossible. For $r > r_B$ the energy levels group into well-defined manifolds, which are approximately spaced by an energy $2B$, corresponding to a quantum of rotational excitation. In the following we focus on this region $r>r_B$, where the adiabatic approximation can be fulfilled.

Figures~\ref{fig:fig4}(b,c) and Figs.~\ref{fig:fig4}(d,e) are
blow-ups of the two lowest-energy manifolds of
Fig.~\ref{fig:fig4}(a) in the region $r > r_B$, for $\vartheta=\pi/2$ and $\vartheta=0$,
respectively. %Different from the zero-field case,
Figure.~\ref{fig:fig4}(b) and Fig.~\ref{fig:fig4}(d) show that the
excited state manifold with one quantum of rotation ($J_1+J_2=1$) is
asymptotically split into two sub-manifolds. This separation
corresponds to the electric-field-induced splitting of the $J_j=1$
manifold of each molecule, and it is thus given by $\hbar \delta = 3
B \beta^2/20$, see caption of Fig.~\ref{fig:fig2}. More importantly,
Figs.~\ref{fig:fig4}(c) and (e) show that the
effective ground-state potential $E_0({\bf r})$ has a very different
character for the cases $\vartheta=\pi/2$ and $\vartheta=0$,
respectively. In fact, in the case $\vartheta=\pi/2$ [Fig.~\ref{fig:fig4}(c)],  corresponding to collisions in the $(z=0)$-plane, the
potential is attractive for $r < r_\star$, while for $r>r_\star$ it turns into repulsive and it decays  at large distances as $1/r^3$, where $r_\star$ is a characteristic length to be defined below.
On the other hand, for $\vartheta=0$ [see Fig.~\ref{fig:fig4}(e)]
the potential is purely attractive, with
dipolar character.
This change in character of the ground-state potential as a function
of $\vartheta$ is captured by the following analytic expression for
$E_{0;0;+}({\bf r})$, as derived in perturbation theory in $V_{\rm
dd}({\bf r})/B$,
\begin{eqnarray}\label{eq:eqVeff3D}
    V_{\rm eff}^{\rm 3D}({\bf r}) \equiv  E_{0;0;+}({\bf
      r})\approx \frac{C_{3}}{r^3}\left(1-3\cos^2\vartheta\right)
+\frac{C_{6}}{r^6}.
  \end{eqnarray}
Here, the constants $C_{3}\approx
d^2\beta^2/9$ and $C_{6}\approx - d^4/6B$ are the dipolar and van-der-Waals coefficients for the ground-state, respectively, and the constant term $2E_{0,0}=-\beta^2 B/3$ due single-particle DC Stark-shifts has been neglected. Equation~\eqref{eq:eqVeff3D} is valid for
$r\gg r_B$ and  $V_{\rm
dd}({\bf r})/B \ll 1$, and it shows that the
potential $V_{\rm eff}^{\rm 3D}({\bf r})$ has a local maximum in the plane $z=r \cos \vartheta = 0$
at the position $r_\star$, defined as
\begin{eqnarray}\label{eq:rstar}
r_\star\equiv
\left(\frac{2|C_{6}|}{C_{3}}\right)^{1/3}\approx\left(\frac{3d^2}{B\beta^2}\right)^{1/3},
\end{eqnarray}
where the dipole-dipole and van-der-Waals interactions become comparable.
The height of this maximum is
\begin{eqnarray}\label{eq:Vstar}
V_\star=\frac{{C_{3}}^2}{4|C_{6}|}\approx \frac{B\beta^4}{54},
\end{eqnarray}
and the curvature of the potential along $z$ [$\partial_z^2 V(r=r_\star,z=0) =
-6C_{3}/r_\star^5 \equiv - m\omega_{\rm c}^2/2$] defines a
characteristic frequency
\begin{eqnarray}\label{eq:omegac}
\omega_{\rm c}\equiv
\left(\frac{12C_{3}}{mr_\star^5}\right)^{1/2},
\end{eqnarray}
to be used below. The latter has a strong dependence $\beta^{8/3}=(d
E_{\rm DC}/B)^{8/3}$ on the applied electric field.
For distances $r \gg r_\star$ the dipole-dipole interaction dominates over the van-der-Waals attractive potential, and $V_{\rm eff}^{\rm 3D}({\bf r})\sim
C_{3}(1-3\cos^2\vartheta)/r^3$, see Ref.~\cite{Buechler07}. Thus, if it were possible to confine the collisional
dynamics to the $(z=0)$-plane with $r \gg r_\star,r_B$, purely repulsive long-range
interactions with a characteristic dipolar spatial dependence $\sim
1/r^3$ could be attained. In the following we analyze the
conditions for realizing this setup, using a strong confinement along $z$
as provided, e.g. by an optical trapping potential.\\

{\em Parabolic confinement} :- The presence of a finite trapping potential of frequency
$\omega_{\perp}$ in the $z$-direction provides for a
position-dependent energy shift of Eq.~\eqref{eq:eqVeff3D}. The new
potential reads
\begin{eqnarray}\label{eq:eqE0r}
  V({\bf r}) %V_{\rm eff}^{\rm 3D}({\bf r}) + \frac{1}{4}m \omega_\perp^2
%  z^2 \nonumber\\
  =  \frac{C_{3}}{r^3}\left(1-3\cos^2\vartheta\right)
  +\frac{C_{6}}{r^6} + \frac{1}{4}m \omega_\perp^2z^2.
  \end{eqnarray}

As noted before, for $z=0$ the repulsive dipole-dipole interaction
dominates over the attractive van-der-Waals at distances $r\gg
r_\star$ given in Eq.~\eqref{eq:rstar}. In addition, for
$\omega_\perp
> 0$ the harmonic potential confines the particle's motion in the
$z$ direction. Thus, the combination of the dipole-dipole
interaction and of the harmonic confinement yields a repulsive
potential which provides for a {\em three-dimensional} barrier separating
the long-distance from the short-distance regime. If the collisional
energy is much smaller than this barrier, the particle's motion is
confined to the long-distance region, where the potential is purely
repulsive.\\ %Below, we explain in some details the conditions for the
%stability of this three-dimensional setup.\\

\begin{figure*}%[htbp]
  \begin{center}
    \includegraphics[width=\textwidth]{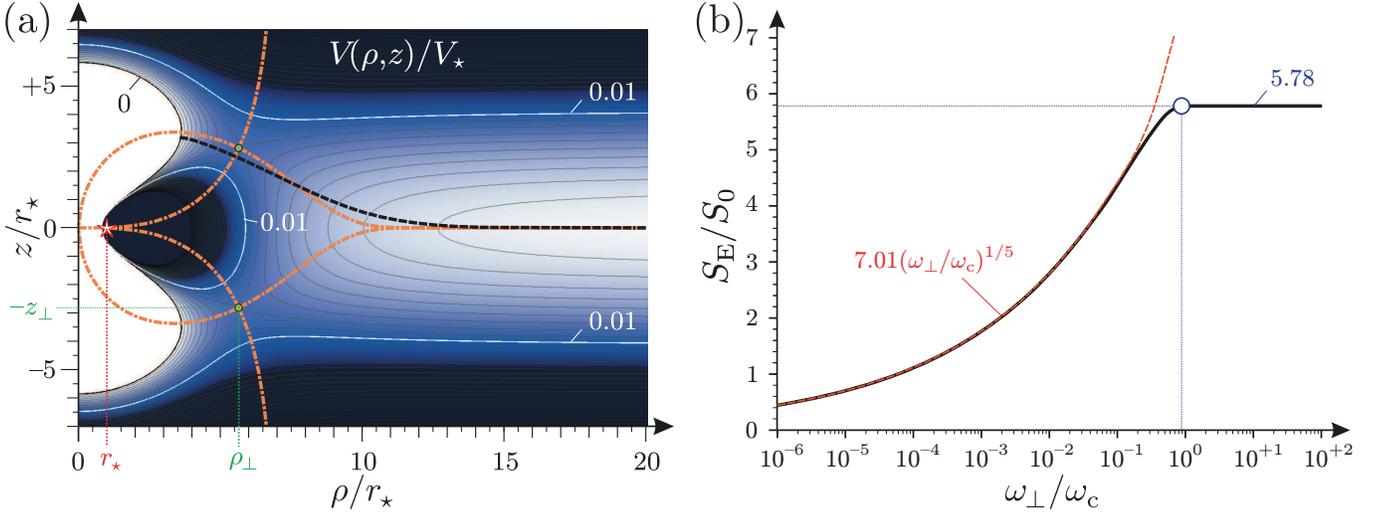}%{figs/Figure_7.eps}%{fig07.eps}
  \end{center}
  \caption{\label{fig:fig5}(a) Contour plot of the effective potential $V(\rho,z)$
  of Eq.~\eqref{eq:eqE0r}, for two polar molecules interacting
  in the presence of a DC field $\beta>0$, and a confining harmonic potential in the
  $z$-direction, with trapping frequency $\omega_\perp=\omega_{\rm
  c}/10$, where $\omega_{\rm c}\equiv (12C_{3}/mr_\star^5)^{1/2}$ of Eq.~\eqref{eq:omegac} and
  $r_\star=(2 |C_{6}|/C_{3})^{1/3}$ of Eq.~\eqref{eq:rstar}. The contour lines are shown for $V(\rho,z)/V_\star \geq 0$, with
    $V_\star=B \beta^4/54$.
  Darker regions represent stronger repulsive interactions. The
    combination of the dipole-dipole interactions induced by the
    DC field and of the harmonic confinement leads to
    realizing a 3D repulsive potential. The
  repulsion due to the dipole-dipole interaction and of the harmonic
  confinement is distinguishable at $z\sim 0$ and $z/r_\star\sim
  \pm 7$, respectively. Two saddle points (circles) located at
  ($\rho_\perp,\pm z_\perp$) separate the long-distance region where
  the potential is repulsive $\sim 1/r^3$ from the attractive
  short-distance region. The gradients of the potential
  are indicated by dash-dotted lines. The thick dashed line
    indicates the instanton solution for the tunneling through the
    potential barrier. (b) The euclidian action $S_{\rm E}$
    as a function of $\omega_\perp/\omega_{\rm c}$ (solid line). For $\omega_\perp<\omega_{\rm c}'\approx
0.88~\omega_{\rm c}$ ($\omega_\perp>\omega_{\rm c}'$) the "bounce"
occurs for $z(0)\neq0$ (within the plane $z(0)=0$), see text. The
point $\omega_{\rm c}'$ is signaled by a circle. For
    $\omega_\perp>\omega_{\rm c}'$ the action is $S_{\rm E}\approx5.78 S_0$, with
    $S_0=\sqrt{m |C_{6}|}/\hbar r_\star^2$,
    which is $\omega_\perp$-{\em independent}, consistent with the
    "bounce" occurring in the ($z=0$)-plane (see text).}
\end{figure*}

Figure~\ref{fig:fig5} is a contour plot of $V({\bf r})$ in units of
$V_\star$, for $\beta>0$ and $\omega_\perp=\omega_{\rm c}/10$, with
${\bf r}\equiv(\rho,z)=r(\sin \vartheta,\cos\vartheta)$ (the angle
$\varphi$ is neglected due to the cylindrical symmetry of the
problem). Darker regions correspond to a stronger repulsive
potential, while the white region at $\rho \approx 0$ is the short-range, attractive part of the potential. The repulsion due to the dipole-dipole and harmonic
potentials is distinguishable at $|z|/r_\star \sim 0 $ and $7$,
respectively. The lesser-dark regions located at $(\rho_\perp,\pm
z_\perp)\equiv
\ell_\perp(\sin\vartheta_\perp,\pm\cos\vartheta_\perp)$ correspond
to the existence of two saddle points positioned in between the maxima of $V({\bf r})$, with $\ell_\perp=(12C_{3}/m\omega_\perp^2)^{1/5}$ and
$\cos\vartheta_\perp=\sqrt{1-(r_\star/\ell_\perp)^3}/\sqrt{5}$, [see circles in Fig.~\ref{fig:fig5}]. These saddle points act as en effective potential barrier separating the attractive part of the potential present at $r< l_{\perp}$ from the region $r \gg \ell_\perp \geq
r_\star,r_B $ where the effective interaction potential
Eq.~\eqref{eq:eqE0r} is purely repulsive. For collisional energies smaller than the height of this barrier the dynamics of the particles can be
reduced to a {\em quasi} two-dimensional (2D) dynamics, by tracing over
the fast particle motion in the $z$-direction. We notice that the existence of two
saddle points at distances $r\sim \ell_\perp$ separating the long-
from the short-distance regimes is a general feature of systems with a comparatively weak transverse trapping
$\omega_{\perp}/\omega_{\rm c}<1$, with $\omega_{\rm c}$ defined in Eq.~\eqref{eq:omegac}. In fact, for a strong transverse trapping
$\omega_\perp{\geq}\omega_{\rm c}$ the two saddle points collapse
into a single one located at $z=0$, and $\rho = \ell_\perp \sim
r_\star$. In this limit the dynamics is purely 2D, with the
particles strictly confined to the $(z=0)$-plane.\\

{\em Collisional stability} :- When an {\em ensemble} of polar molecules is considered, inelastic
collisions and three body recombination may lead the system to a
potential instability, associated with the attractive character of
the dipole-dipole
interaction~\cite{Rot1,Rot3,Rot4,Rot5,Rot6,Rot7,Rot8,Rot9,Rot10}. In our
discussion, this instability is associated with the population of
the short-distance region $r < \ell_\perp$, which can be efficiently
suppressed for strong dipole-dipole interactions and transverse confinement. In fact, for collisional energies smaller than the
potential barrier $V(\rho_\perp,\pm z_\perp)$ the particles are
mostly confined to the long-distance regime, where they scatter
elastically. That is, when a cold ensemble of molecules is
considered the barrier provides for the stability of the system by
``shielding'' the short-distance attractive part of the two-body
potential. In this limit, residual losses are due to the tunneling
through the potential barrier at a rate $\Gamma$, which can be
efficiently suppressed for reasonable values of $\beta$ and
$\omega_\perp$, as shown below.\\

The tunneling rate $\Gamma = \Gamma_0 e^{-S_{\rm E}/\hbar}$ through the barrier $V(\rho_\perp,\pm z_\perp)$ can be calculated using a semi-classical/instanton
approach~\cite{Coleman77}. The euclidian action $S_{\rm E}$, which
is responsible for the exponential suppression of the tunneling, is
plotted in Fig.~\ref{fig:fig5} as a function of
$\omega_\perp/\omega_{\rm c}$, in units of
$S_0=\sqrt{m|C_{6}|}/r_\star^2$. %=(2Bd^4m^3\beta^8/3^7)^{1/6}$.
The figure shows that $S_{\rm E}$ has different behaviors for
$\omega_\perp\ll \omega_{\rm c}$ and $\omega_\perp\gg\omega_{\rm
c}$. In fact, for $\omega_\perp\ll \omega_{\rm c}$ the action
increases with increasing $\omega_\perp$ as $S_{\rm E}\approx
7.01S_0(\omega_\perp/\omega_{\rm c})^{1/5} =
%5.86(C_{3;0}^2m^3\omega_\perp/8)^{1/5}=
1.43\hbar(\ell_\perp/a_\perp)^{2}$ (dotted line), which depends on
%the $C_{3;0}$-coefficient of the dipole-dipole interaction and the
the confinement along $z$, {\em via} $a_\perp=(\hbar/m
\omega_\perp)^{1/2}$. On the other hand, for
$\omega_\perp\gg\omega_{\rm c}$ it reads $S_{\rm E}\approx
5.78~S_0$, which is $\omega_\perp$-independent. The transition
between the two different regimes mirrors the change in the nature
of the underlying potential $V({\bf r})$. In particular, for $\omega_\perp
\gtrsim \omega_{\rm c}$ the dynamics is strictly confined to the plane $z=0$ and
thus it becomes independent of $\omega_{\perp}$. The constant $\Gamma_0$ is
related to the quantum fluctuations around the semiclassical
trajectory, and its value is strongly system-dependent. For the
crystalline phase of Ref.~\cite{Buechler07}, it is the collisional
"attempt frequency", proportional to the characteristic phonon
frequency $\Gamma_0\sim \sqrt{C_{3}/m a^5}$, with $a$ the mean
interparticle distance.

In the limit of strong interactions and tight transverse confinement
$\Gamma$ rapidly tends to zero. We illustrate this for the example
of SrO, which has a permanent dipole-moment of $d\approx 8.9~{\rm
Debye}$ and mass $m=104~{\rm amu}$. Then, for a tight transverse
optical lattice with harmonic oscillator frequency
$\omega_\perp=2\pi\times150{\rm kHz}$ and for a DC-field
$\beta=dE_{\rm DC}/B=1/3$ we have
$(C_{3}^2m^3\omega_\perp/8\hbar^5)^{1/5}\approx 3.39$ and we obtain
$\Gamma/\Gamma_0\approx e^{-5.86\times3.39}\approx 2\times10^{-9}$.
Even for a DC field as weak as $\beta=1/6$ we still
obtain a suppression by five order of magnitudes, as
$\Gamma/\Gamma_0\approx e^{-5.86\times1.94}\approx 10^{-5}$.
This calculation confirms that a collisionally stable setup for polar molecules in the strongly interacting regime can be realized by combining the strong dipole-dipole interactions with a tight transverse confinement.\\

{\em Effective 2D interaction}:- The effective two-dimensional interaction potential is obtained by
integrating out the fast particle motion in the transverse direction
$z$. For $r > \ell_\perp \gg a_\perp$, the two-particle
eigenfunctions in the $z$-direction approximately factorize into
products of single-particle harmonic oscillator wave-functions
$\psi_{k_1}(z_1)\psi_{k_2}(z_2)$, and in first order perturbation
theory in $V_{\rm eff}^{\rm 2D}/\hbar\omega_\perp$ the effective 2D
interaction potential $V_{\rm eff}^{\rm 2D}$ reads
\begin{eqnarray}\label{eq:eff2dpotential}
V_{\rm eff}^{\rm 2D}(\boldrho) \approx %\int dz_1dz_2
  %\psi_0(z_1)^2\psi_0(z_2)^2 V_{\rm eff}^{\rm 3D}({\bf r})\nonumber\\
  \frac{1}{\sqrt{2\pi}a_\perp}\int dz e^{-z^2/2a_\perp^2}V_{\rm eff}^{\rm 3D}({\bf
  r}).
\end{eqnarray}
For large separations $\rho\gg \ell_\perp$ the 2D potential reduces
to
\[
V_{\rm eff}^{\rm 2D}(\boldrho)=\frac{C_{3}}{\rho^3},
\]
\\
which is a purely repulsive 2D interaction potential. The derivation
of $V_{\rm eff}^{\rm 2D}(\boldrho)$ is one of the central results of
this section. We show below (Sect.~\ref{sec:secApplications}) that the use
of this interaction potential leads to the realization of
interesting many-body phases, in the context of condensed matter
applications using cold molecular quantum gases.

 \begin{figure*}[htb]
 \begin{center}
 \includegraphics[width=\textwidth]{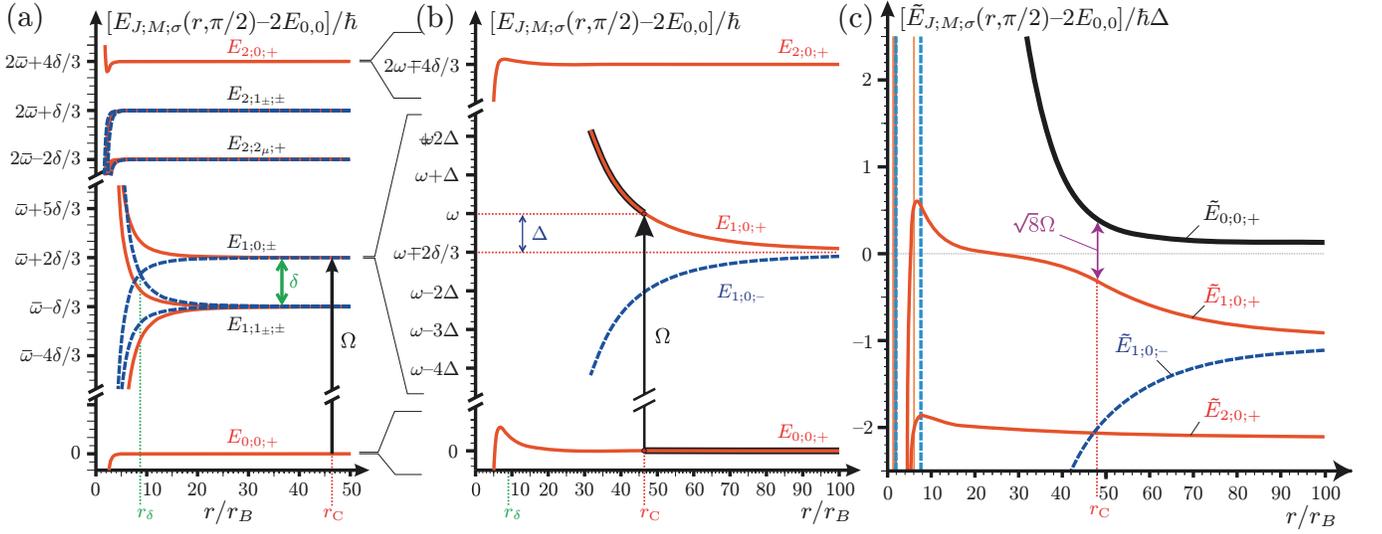}%{figs/Figure_8.eps}%{fig12.eps}
  \caption{\label{fig:fig11}(a) Schematic representation of the effects
    of a DC and an AC microwave fields on the interaction of two molecules.
    The solid and dashed lines are the bare
    potentials $E_n({\bf r})\equiv E_{J;M;\sigma}(r,\vartheta)$ of
    Sec.~\ref{sec:secDC} with $\vartheta=\pi/2$
    for interactions in the presence of the DC field only,
    for the symmetric ($\sigma=+$) and antisymmetric ($\sigma=-$)
    states,
    respectively. The DC field induces a splitting $\hbar\delta$
    of the first-excited manifold of the two-particle spectrum.
    A microwave-field of frequency $\omega=\overline{\omega}+2\delta/3+\Delta$ is blue
    detuned by $\Delta>0$ from the single-particle rotational
    resonance. The dipole-dipole interaction further splits the excited-state
    manifold, making the detuning space-dependent. Eventually, the
    combined energy of the bare ground-state potential $E_{0;0;+}({\bf r})$
    and of an AC photon (black arrow) becomes degenerate with the energy of the
    bare symmetric $E_{1;0;+}(r,\pi/2)$. The resonant point $r_{\rm C} =
(d^2/3 \hbar\Delta)^{1/3}$ occurs at $r\approx46~r_B$. %A second
%    resonant Condon point occurs at (much) shorter distances
%    $r_{\rm C'}\lesssim r_{\delta}=(d^2/\hbar\delta)^{1/3}$
%    with an anti-symmetric potential (not shown).
    (b) Blow-up of the
    potentials of panel (a) with $M=0$.
    The {\em dressed} ground-state potential is sketched by a thick solid
    line. (c) The four potentials of panel (b) in the field-dressed picture. The dressed ground-state potential $\tilde{E}_{0;0;+}(r,\pi/2)$ has the largest energy and is indicated by a thick solid line.}
 \end{center}
 \end{figure*}%

\subsubsection{{\em {\bf Designing {\em ad-hoc} potentials with AC-fields}}}\label{sec:secAC}

Above we have shown how to design 2D effective groundstate interactions which are purely repulsive and decay as $\sim 1/r^3$. The use of one or several non interfering AC fields allows to engineer more complicated interactions by combining the spatial texture of the adiabatic groundstate potential of the two-particle spectrum with that of selected excited potentials, in a dressed picture. This mixing of ground and excited-state potentials is favored by the dipole-dipole interactions which split the degeneracy of the excited-state manifolds of the two-particle spectrum and render the state-selectivity of the AC fields {\em space-dependent}, as explained below.
In combination with a strong optical confinement, and due to the long lifetimes of the excited rotational states \cite{Kotochigova:04}, this allows for the realization of collisionally stable setups for molecules in the strongly interacting regime.\\

We exemplify the situation above by considering the case of a single AC field ${\bf E}_{\rm AC}(t) = E_{\rm
AC}e^{-i\omega t}{\bf
  e}_q + {\rm c.c.}$ which is added to the configuration of
Fig.~\ref{fig:fig4} (interactions in the presence of a static
electric field ${\bf E}_{\rm DC}=\beta B {\bf e}_z$). The
field's polarization is chosen to be linear ($q=0$) and the frequency $\omega$ is
blue-detuned from the ($\ket{\phi_{0,0}} \rightarrow
\ket{\phi_{1,0}}$)-transition of the single-particle spectrum by an
amount $\Delta=\omega-2B/\hbar>0$.
The effects of the AC-field on the two-particle scattering can be
summarized as: (a) Inducing {\em oscillating} dipole-moments in each
molecule, which determine long-range dipole-dipole interactions [in
addition to those determined by the static field ${\bf E}_{\rm
DC}$], whose sign and
angular dependence are given by the polarization $q$; %and the
%orientation in space, ${\bf e}_r$;
(b) Inducing a coupling of the ground and excited state manifolds of
the {\em two-particle} spectrum at a resonant (Condon) point $r_{\rm
C}=(d^2/3h \Delta)^{1/3}$, where the dipole-dipole interaction
becomes comparable to the detuning $\Delta$. This coupling is
responsible for an avoided crossing of the {\em field-dressed}
energy levels at $r_{\rm C}$, whose properties depend crucially on
the polarization $q$. This fact is at the core of the engineering of
interaction potentials, in that the 3D effective {\it dressed
adiabatic} ground-state interaction potential inherits the character
of the bare ground and excited potentials for $r \gg r_{\rm C}$ and
$r \ll r_{\rm C}$, respectively.

The setup above is illustrated in Fig.~\ref{fig:fig11}(a) and
(b), where the continuous
and dashed lines are the bare (${\bf E}_{\rm AC} =0$) symmetric and
anti-symmetric potentials $E_{J;M;\sigma}({\bf r})$ of
Fig.~\ref{fig:fig4}, respectively, and the presence of the AC-field
is signaled by a black arrow at the resonant Condon point $r_{\rm
C}$. The presence of the weak DC field splits asymptotically the ($J=1$)-manifold by an amount $\hbar \delta$ as in Fig.~\ref{fig:fig4}(b), allowing for a simple fulfillment of the adiabatic approximation in the excited-state manifold for distances $r\gg r_{\delta}=(d^2/\hbar \delta)^{1/3}$. In fact, the energy of the $E_{1;0;+}({\bf r})$ potential becomes degenerate with the energy of other bare symmetric potentials only at distances $r \ll r_\delta$. In addition, we notice that the presence of the splitting $\hbar \delta$ also shifts the level crossing with antisymmetric states %$E_{1;1_+;-}({\bf r})$
to small distances $r \ll r_\delta$.

For distances $r\gg r_{\delta}=(d^2/\hbar \delta)^{1/3}$, we are allowed to consider only the four states of
Fig.~\ref{fig:fig11}(b), since all the other potentials of the
($J=1$)- and ($J=2$)-manifolds are far detuned by an amount which is (at least) of order $\delta \gg \Delta$ and they are not coupled by the AC-field to
the bare ground state $E_{0,0;+}({\bf r})$, due to the choice of field's polarization.
Figure~\ref{fig:fig11}(b) shows that the splitting induced by the
dipole-dipole interaction in the ($J=1$)-manifold renders the
detuning $\Delta$ position-dependent, so that at $r_{\rm C}$ the
energy of the bare ground-state and that of the symmetric bare
excited state become degenerate. The resulting dressed ground-state
potential is sketched in Fig.~\ref{fig:fig11}(b) (thick
black line) and it roughly corresponds to the bare $E_{0,0;+}({\bf r})$ and $E_{1,0;+}({\bf r})$ potentials for $r> r_C$ and $r<r_C$, respectively.
Accordingly, Fig.~\ref{fig:fig11}(c) shows that the dressed groundstate potential $\tilde{E}_{0;0;+}({\bf r})$, which has the {\em highest energy}, turns from weakly to strongly repulsive for $r\gg r_{\rm C}$ and $r\ll r_{\rm C}$,
respectively. This change in the character of the ground-state
interaction potential corresponds to the design of a "step-like"
interaction. This example shows that the 3D ground-state
interaction for two molecules can be strongly modified by
the combined use of AC and DC fields, which is the central result of
this section. More complicated potentials can be engineered using
multiple AC fields
and different polarizations.\\

Analogous to the case (${\bf E}_{\rm AC}=0$) of
Sect.~\ref{sec:secDC}, the interaction potential of
Fig.~\ref{fig:fig11} is actually repulsive along certain directions
(e.g. $\theta=\pi/2$, as shown in the figure), while it turns into
attractive along others (e.g. $\theta = 0$, not shown). As for the
(${\bf E}_{\rm AC}=0$)-case of Sect.~\ref{sec:secDC}, when more than two particles are
considered this attraction can lead to many-body instabilities.
Moreover, here the dressed potential $\tilde{E}_{0;0;+}({\bf r})$ of Fig.~\ref{fig:fig11}(c) is {\it not} the lowest-energy
potential, which in general can introduce additional loss channels. %in the {\em
%two-particle} collision.
The latter correspond to diabatic couplings to symmetric
states for particles approaching distances $r \lesssim r_{\rm
C}$, and are therefore present even in the simple {\em two-particle} collisional process, and to couplings to anti-symmetric states, which can be induced e.g. by
three-body collisions or by non-compensated tensor-shifts for two optically-trapped particles. The presence of all of these loss channels may render impractical
the realization of collisionally stable setups for strongly
interacting molecular gases (although  Ref.~\cite{Gorshkov08} for a solution involving the use of a {\em circularly} polarized AC field). However, we have seen above that for the setup of Fig.~\ref{fig:fig11} the presence of the static field shifts the various resonance points with the potentials which are responsible for these loss channels {\em in the} ($\vartheta=\pi/2$)-{\em plane} ($z=0$) to distances $r \ll r_C$. This suggests that by confining the particles motion {\em to the plane $z=0$} by using a strong optical transverse confinement analogous to that of Sect.~\ref{sec:secDC} it is possible to realize collisionally stable setups in the region $r > r_C$ . This scheme has been shown to work in Ref.~\cite{Micheli07} and thus the main message here is that a judicious combination of the dipole-dipole interactions and of the optical confinement can act as an effective "shield" of the region $r<r_C$ where losses occur and thus the collisional setup can be made stable. The use of the step-like potential above and of other engineered potentials can lead to the realization of interesting phases for an ensemble of
polar molecules in the {\em strongly interacting} regime
\cite{Buechler07,Micheli07}.

\section{Many-body physics with cold polar molecules}\label{sec:secApplications}

\subsection{2D Self-Assembled crystals}
\begin{figure*}%[htb]
\begin{center}
\includegraphics[width=\textwidth]{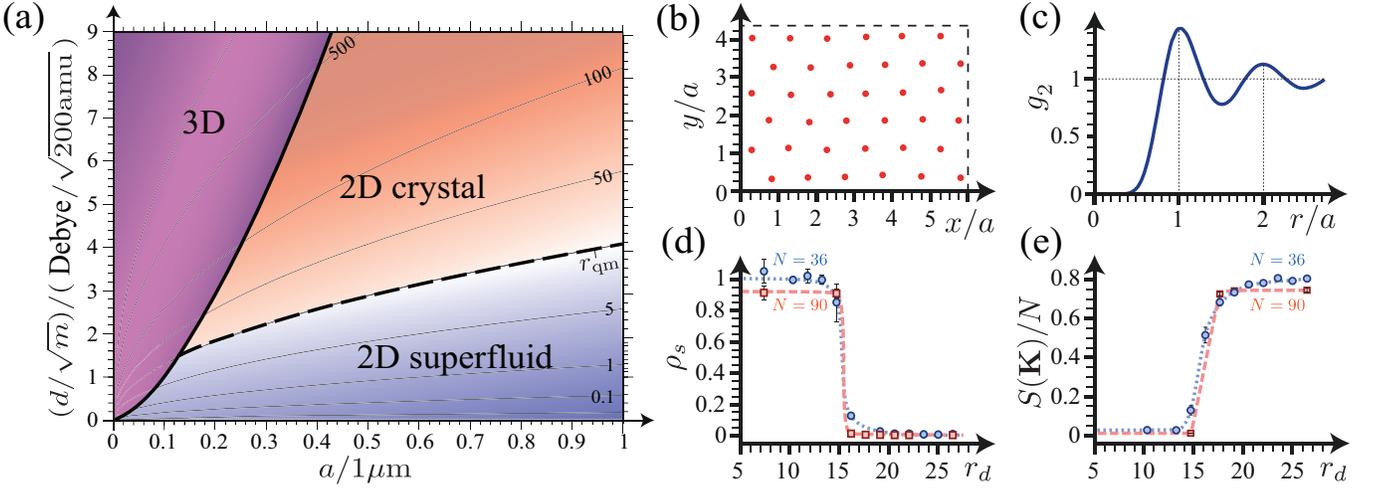}%{Fig3_revised.eps}
\end{center}
\caption{(a) Quantum phases of 2D dipoles: Contour plot of the interaction strength $r_d=D m /\hbar^2 a$ as a function of the dipole moment $d$ (in Debye) and of the interparticle distance $a$ (in $\mu$m), with $m$ the mass of a molecule (in atomic units $200 \times$amu). The regions of stability of the 2D superfluid and crystalline phases where $\hbar \omega_\perp > D/a^3$ are indicated, with $\omega_\perp=2 \pi \times 150$kHz the frequency of the transverse confinement. (b) PIMC-snapshot of
the mean particle positions in the crystalline phase for $N=36$ at
$r_{d}\approx26.5$. (c) Density-density (angle-averaged) correlation
function $g_{2}(r)$, for $N=36$ at $r_{d}\sim11.8$. (d) Superfluid
density $\rho_{s}$ and (e) static structure factor $S({\bf K})/N$ as
a function of $r_{d}$, for $N=36$ (circles) and $N=90$ (squares).}
\label{fig3}
\end{figure*}

The above discussion of the intermolecular potentials and of the
stability of collisional setups in reduced dimensionality provides
the microscopic justification for studying an ensemble of polar
molecules in 2D interacting {\it via} (modified) dipole-dipole
potentials. At low temperatures $T<\hbar \omega_{\perp}$, the
general {\it many body Hamiltonian} has the form of Eq.~\eqref{hamilton1}.
%WE HAD THIS HAMILTONIAN BEFORE IN SEC II AND I SUGGEST THAT WE GIVE REFERENCE AND DO NOT WRITE IT AGAIN. IN FACT MOST OF THE FOLLOWING PARAGRAPHS BELOW ARE IDENTICAL TO WHAT WE HAVE IN SEC II. IN THE PRESENT FORM TOO LITTLE MAY BE LEFT OF THIS SECTION - SEE MY COMMENTS IN SEC II
%\begin{equation}
% H = \sum_{i} \frac{{\bf p}_{i}^{2}}{2 m} + \sum_{i < j}
% V_{\rm eff}^{\rm 2D}(\boldrho_{ij}). \label{hamilton}
%% V_{\rm eff}^{\rm 2D}({\bf R}_{i}- {\bf R}_{j}). \label{hamilton}
%\end{equation}
%
%The first term accounts for the kinetic energy within the
%$x$,$y$-plane, while the second term  denotes the effective
%interaction potential, with $\boldrho_{ij}\equiv(x_j-x_i,y_j-y_i)$ a
%two-dimensional vector. \\
As an example of the possibilities offered  by potential engineering
to realize novel many-body quantum phases, we here focus on bosonic particles interacting via the effective potential
\begin{eqnarray}
V_{\rm eff}^{\rm 2D}(\boldrho)= D/\rho^3 \label{eq:eqIntPot}
\end{eqnarray}
derived in Sect.~\ref{sec:secDC}. This is the simplest attainable
interaction potential, and thus the one which is experimentally most
interesting in the short term. However, despite the simplicity of
the interaction potential, the Hamiltonian Eq.~(\ref{hamilton1})
gives rise to novel quantum phenomena, which have not been accessed
so far in the context of cold neutral atoms and molecules. In
particular, by means of Path Integral Monte-Carlo simulations
(PIMC), in Ref.~\cite{Buechler07} we show the appearance of a {\em
self-assembled} crystalline phase and an associated quantum melting
transition into a superfluid as a function of the interaction
strength $r_d$. As explained in Sect.~\ref{sec:secIntro2}, the latter is the ratio
\begin{eqnarray}
r_d=\frac{D m}{\hbar^2 a}\label{eq:eqrd}
\end{eqnarray}
between the interaction energy $D/a^3$ and the kinetic energy
$\hbar^2/m a^2$ at the mean interparticle distance $a$, with $m$ the
mass of a molecule.\\

In Fig.~\ref{figs:fig1000}(a) a tentative phase diagram is sketched for the
two-dimensional system of bosonic dipoles. In the limit of
weak interactions $r_{d}<1$, the ground state is a superfluid (SF)
with a finite (quasi) condensate. The SF is characterized by
a superfluid fraction $\rho_{s}(T)$, which depends on temperature $T$, with $%
\rho_{s}(T=0)=1$. Since we consider a 2D setup, a Berezinskii--Kosterlitz--Thouless transition %\cite{KosterlitzThoulessJPhysC6-1181-1973}
towards a normal fluid is expected to occur at a finite temperature
$T_{\mathrm{\scriptscriptstyle KT}}= \pi\rho_{s}\hbar^{2}n/2 m$. In
the opposite limit of strong interactions  $r_{d} \gg1$ the polar molecules
are in a crystalline phase for temperatures $T< T_{m}$ with $
T_{m}\approx0.09 D/a^{3}\simeq 0.018r_d E_{{\rm R,c}}$, while for larger temperature the crystal melts into a normal fluid via a first-order (classical) phase transition. The critical value $T_m$ for this melting transition has been obtained via molecular dynamics simulations in the context of interfacial colloidal crystals in Ref.~\cite{Kalia}. Here, $E_{{\rm R,c}}\equiv \pi^2 \hbar^2/2 m a^2$ is the crystal recoil
energy, typically a few to tens of kHz.
The configuration with minimal energy is thus a triangular lattice with spacing $a_{%
\mathrm{\scriptscriptstyle L}} = (4/3)^{1/4} a$. Excitations of the
crystal are acoustic phonons with Hamiltonian
\begin{eqnarray}H_{{\rm c}}=\sum_{q}\hbar\omega_{ q}a_{
q}^{\dagger}a_{q},\label{eq:eqPhonons}
\end{eqnarray}
where $a_{q}$ destroys a phonon of quasimomentum ${\bf q}$ in the
mode $\lambda$. The characteristic Debye frequency is
$\hbar\omega_{{\rm D}}\sim1.6\sqrt{r_{d}}E_{{\rm R,c}}$.
At $T=0$ the static structure factor
$S$ diverges at a reciprocal lattice vector $\mathbf{K}$, and thus $S(\mathbf{K}%
)/N$ acts as an order parameter for the crystalline phase.

In Ref.\cite{Buechler07} we investigated the intermediate strongly
interacting regime with $r_{d}\gtrsim1$, and we determined the
critical interaction strength $r_{\mathrm{ \scriptscriptstyle QM}}$
for the quantum phase transition between the superfluid and the
crystal. In our analysis we used a recently developed PIMC-code
based on the Worm algorithm \cite{Boninsegni06}, which is an {\em
exact} Monte-Carlo method for the determination of thermodynamic
quantities in continuous space at small finite temperature. In
Fig.~\ref{fig3}(d-e), the order parameters $\rho_{s}$ and
$S(\mathbf{K})/N $ are shown at a small temperature $T= 0.014
D/a^{3}$ for different interaction strengths $r_{d}$ and
particle numbers $N=36,90$. We find that $\rho_{s}$ exhibits a
sudden drop to zero for $r_{d}\approx 15$, while at the same
position $S(\mathbf{K})$ strongly increases. In addition, during the Monte-Carlo simulations we observed that in a few occasions $\rho_s$ suddenly jumped from 0 to 1, and then returned to 0, in the interval $r_d\approx 15-20$, which
suggests a competition between the superfluid and crystalline
phases. These results
indicate a superfluid to crystal phase transition at
\begin{eqnarray}
r_{\mathrm{\scriptscriptstyle QM }}= 18\pm4.
\end{eqnarray}
The step-like behavior of $\rho_{s}$ and $S({\bf K})/N$
is consistent with a first order phase transition, a result which
has been confirmed in Refs.~\cite{Astrakharchik07,Mora07}.
We notice that the superfluid with $r_{d}\sim 1$ is strongly interacting,
and in particular the density-density correlation function is quenched
on lengths $R<a$, see Fig.~\ref{fig3}(c). This observation is
consistent with the validity condition of the effective 2D
interaction potential Eq.~\eqref{eq:eqIntPot}, that two particles
never approach each other at distances smaller than $l_{\perp}$.\\

Having determined the low-temperature phase-diagram, the remaining
question is whether these phases, and in particular the crystalline
phase emerging at strong dipole-dipole interacions, are in fact
accessible with polar molecules. This question is addressed in
Fig.~\ref{fig3}(a), which is a contour plot of the interaction
strength $r_d$ as a function of the induced dipole moment $d=\sqrt{D}$ (in units of Debye) and of the mean interparticle
distance $a$ (in $\mu$m). The dimensionless quantity $\sqrt{m/200 amu}$ depends on the mass $m$ of the molecules (in
atomic units), and it is of order one for characteristic molecules
like SrO or RbCs. In the figure, stable 2D configurations for the
molecules exist in the parameter region where the transverse
(optical) trapping frequency $\omega = 2 \pi \times 150$Hz
exceeds the dipole-dipole interaction ($\hbar \omega > D/a^3$), such that $l_{\perp}=(12 D/m
\omega_{\perp}^2)^{1/5} < a$, consistent with the stability
discussion of Sect.~\ref{sec:secDC} [notice that $l_{\perp} \sim (D/\hbar \omega_\perp)^{1/3}$ for realistic parameters]. The figure shows that for a
given induced dipole $d$ the ground-state of an ensemble of
polar molecules is a crystal for mean interparticle distances
$l_{\perp}\lesssim a \lesssim a_{\textrm{max}}$, where
$a_{\textrm{max}}\equiv d^2 m/\hbar^2 r_{\rm QM}$
corresponds to the distance at which the crystal melts into a
superfluid. For SrO (RbCS) molecules with permanent dipole moment
$d=8.9$D ($d=1.25$D), $a_{\textrm{min}}\sim 200nm$($100$nm), while
$a_{\textrm{max}}$ can be several $\mu$m. Since for large enough
interactions the melting temperature $T_{\rm m}$ can be of order of
several $\mu$K, the self-assembled crystalline phase should be
accessible for reasonable experimental parameters using cold polar
molecules.

For what concerns the observability of the zero-temperature phases,
Bragg scattering with optical light allows for probing   the
crystalline phase, while the detection of vortices can be used as a
definitive signature of superfluidity. We notice that the 2D (quasi)
condensate involves a fraction of the total density only, and
therefore we expect only small coherence peaks in a time of flight
experiment. \\

Finally, we notice that by adding an additional {\em in-plane}
optical confinement, it is possible to realize strongly interacting
1D phases which are analogous to the 2D crystals discussed above
\cite{Ark1,all13,all14}. For large enough
interactions $r \gg 1$, the phonon frequencies have the simple form
$\hbar\omega_{q}=(2/\pi^{2})\left[12r_{d}f_{q}\right]^{1/2}E_{{\rm
R,c}}$, with $f_{q}=\sum_{j>0}4\sin(qaj/2)^{2}/j^{5}$. The Debye
frequency is $\hbar\omega_{{\rm
D}}\equiv\hbar\omega_{\pi/a}\sim1.4\sqrt{r_{d}}E_{{\rm R,c}}$, while
the classical melting temperature can be estimated to be of the
order of $T_m\simeq 0.2 r_d E_{{\rm R,c}}/k_B $, see Ref.~\cite{all14}.

\subsection{Floating lattices of dipoles}

An interesting possibility offered by the realization of the self-assembled crystals discussed above is to utilize them as floating mesoscopic lattice potentials to trap extra-particles, which can be atoms or polar molecules of a different species. We show below that within an experimentally accessible parameter regime extended Hubbard models with tunable long-range phonon-mediated interactions describe the
effective dynamics of the extra-particles dressed by the lattice phonons.
\begin{figure}[t!]%[htbp!]
\begin{centering}
\includegraphics[width=1\columnwidth]{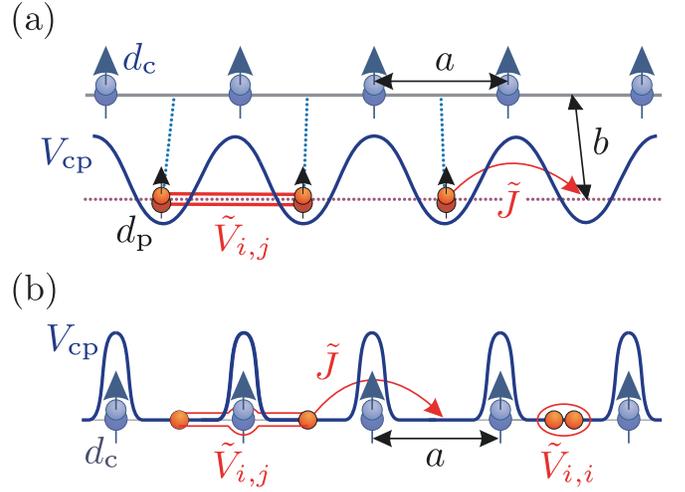}%{fig1}
\par\end{centering}
\caption{\label{figs:fig1}A dipolar crystal of polar molecules provides a periodic lattice $V_{{\rm cp}}$ for
extra atoms or molecules giving rise to a lattice model with hopping
$\tilde{J}$ and long-range interactions $\tilde{V}_{i,j}$ (see
text and Fig.~\ref{fig2}). (a) A 1D dipolar crystal with
lattice spacing $a$ provides a periodic potential for a second
molecular species moving in a parallel tube at distance $b$
(Configuration 1). (b) 1D setup with atoms scattering from the
dipolar lattice (\textit{\emph{Configuration 2}}).}
\end{figure}

The setups that we have in mind are shown in Figs.~\ref{fig2} and~\ref{figs:fig1}(a-b)
where extra particles confined to a 2D crystal plane
or a 1D tube
scatter from the periodic lattice potential $\sum_{j}V_{{\rm
cp}}(\mathbf{R}_{j}-\mathbf{r})$. Here, $\mathbf{r}$ and
$\mathbf{R}_{j}={\bf \mathbf{R}}_{j}^{0}+{\bf \mathbf{u}}_{j}$ are
the coordinates of the particle and crystal molecule $j$,
respectively, with $\mathbf{R}_{j}^{0}$ the equilibrium positions
and $\mathbf{u}_{j}$ small displacements. For particles being
molecules, this potential is given by the repulsive dipole-dipole
interaction $V_{{\rm cp}}({\bf R}_{j}-{\bf r})=d_{{\rm p}}d_{{\rm
c}}/\mid\mathbf{R}_{j}-\mathbf{r}\mid^{3}$ with $d_{{\rm p}}\ll
d_{{\rm c}}$ the induced dipole moment, and for atoms we assume that
the interaction is modeled by a short range pseudopotential
proportional to an elastic scattering length $a_{{\rm cp}}$. In
addition, extra molecules and atoms will interact according to
dipolar, or short range interactions, respectively.

We are interested in a situation where the extra particles in the
lattice are described by a single band Hubbard Hamiltonian coupled
to the acoustic phonons of the lattice \cite{Mahan}

\begin{eqnarray}
H & = & -J\sum_{<i,j>}c_{i}^{\dagger}c_{j}
+\tfrac{1}{2}\sum_{i,j}V_{ij}c_{i}^{\dagger}c_{j}^{\dagger}c_{j}c_{i}\nonumber \\
 & + & \sum_{q,j}M_{ q}e^{i{\bf q}\cdot{\bf R}_{j}^{0}}c_{j}^{\dagger}c_{j}(a_{q}
 +a_{-q}^{\dagger})+H_{{\rm c}}.\label{eq:eqSmallPolaron}\end{eqnarray}
 The first and second terms describe the nearest neighbor hopping of the extra
 particles
with hopping amplitudes $J$, and interactions $V$, computed for each
microscopic model by band-structure calculations for ${\bf u}_j=0$,
respectively. The operators $c_{i}$ ($c_{i}^{\dagger}$) are
destruction (creation) operators of the particles. The third term is
the phonon coupling obtained in lowest order in the displacement
%\begin{eqnarray}
\[{\bf u}_{j}=i\sum_{q}(\hbar/2m_{{\rm c}}N\omega_{ q})^{1/2}\xi_{
q}(a_{q}+a_{- q}^{\dagger})e^{i{\bf q}\cdot{\bf
R}_{j}^{0}},\]%\nolabel\\
%\end{eqnarray}
with
%\begin{eqnarray}
\[M_{q}=\bar{V}_{{\bf q}}{\bf q}\cdot\xi_{ q}(\hbar/2Nm_{{\rm
c}}\omega_{ q})^{1/2}\beta_{{\bf q}}. \] Here, $\xi_{ q}$ and $N$
are the phonon polarization and the number of lattice molecules,
respectively, while $\bar{V}_{{\bf q}}$ is the Fourier transform of
the particle-crystal interaction $V_{\rm cp}$, and $\beta_{{\bf
q}}=\int d\mathbf{r}|w_{0}({\bf \mathbf{r}})|^{2}e^{i\mathbf{qr}}$,
with $w_{0}({\bf \mathbf{r}})$ the Wannier function of the lowest
Bloch band \cite{Mahan}. The validity of the single band Hubbard
model requires $J,V<\Delta$, and temperatures $k_{B}T<\Delta$ with
$\Delta$ the separation to the first excited Bloch band.\\

The Hubbard parameters of Eq.~\eqref{eq:eqSmallPolaron} are of the
order of magnitude of the recoil energy, $J,V\sim E_{{\rm R,c}}$,
and thus (much) smaller than the Debye frequency
$\hbar\omega_{D}\sim E_{{\rm R,c}}\sqrt{r_{d}}$, for $r_d\gg1$
\cite{Albus}. This separation of time scales
$J,V$$\ll\hbar\omega_{D}$, combined with the fact that the coupling
to phonons is dominated by high frequencies $\hbar\omega>J,V$ (see
the discussion of $M_{ q}$ below) is reminiscent of polarons as
particles dressed by (optical) phonons, where the dynamics is given
by coherent and incoherent hopping on a lattice
\cite{Mahan,Alexandrov}. This physical picture is brought out in a
master equation treatment within a strong coupling perturbation
theory. The starting point is a Lang-Firsov transformation of the
Hamiltonian $H\rightarrow \mathcal{S}H\mathcal{S}^{\dagger}$ with a
density-dependent displacement \[\mathcal{S}=\exp\left[-\sum_{ q,j}\frac{M_{
q}}{\hbar\omega_{
q}}e^{i\mathbf{q}\mathbf{R}_{j}^{0}}c_{j}^{\dagger}c_{j}(a_{ q}-a_{-
q}^{\dagger})\right].\] This eliminates the phonon coupling in the second
line of Eq.~\eqref{eq:eqSmallPolaron} in favor of a transformed
kinetic energy term \[-J\sum_{
<i,j>}c_{i}^{\dagger}c_{j}X_{i}^{\dagger}X_{j},\] where the
displacement operators \[X_{j}=\exp\left[\sum_{ q}\frac{M_{
q}}{\hbar\omega_{ q}}e^{i\mathbf{q}\mathbf{R}_{j}^{0}}(a_{ q}-a_{-
q}^{\dagger})\right]\] can be interpreted as a
lattice recoil of the dressed particles in a hopping process. In
addition, the bare interactions are renormalized according to
\[\tilde{V}_{ij}=V_{ij}+V_{ij}^{(1)},\] with $V_{ij}^{(1)}=-
2\sum_{q}\cos({\bf q}({\bf R}_{i}^{0}-{\bf R}_{j}^{0}))M_{
q}^{2}/\hbar\omega_{ q}$, that is, the phonon couplings induce and
modify off-site interactions. The onsite interaction is given by
$\tilde{V}_{j,j}=V_{j,j}-2E_{p}$ with \[E_{p}=\sum_{ q}\frac{M_{
q}^{2}}{\hbar\omega_{ q}}\] the \textit{polaron} self-energy or
\textit{polaron shift}. For $J=0$ the new Hamiltonian is diagonal
and describes interacting polarons and independent phonons. The
latter are vibrations of the lattice molecules around new
equilibrium positions with unchanged frequencies. A stable crystal
requires the variance of the displacements $\Delta u$ around these
new equilibrium positions to be small compared to $a$.

A Born-Markov approximation with the phonons a finite temperature
heatbath with $J,V\ll \hbar\omega_D$ (see above), and the
transformed kinetic energy \[-J \sum_{<i,j>}
c_i^{\dagger}c_j(X_{i}^{\dagger}X_{j}-\langle\langle
X_{i}^{\dagger}X_{j}\rangle\rangle)\] as the system-bath interaction
with $\langle\langle X_{i}^{\dagger}X_{j}\rangle\rangle$ the
equilibrium bath average, provides the master equation for the
reduced density operator of the dressed particles $\rho_t$ in
Lindblad form \cite{Car}
\begin{eqnarray}
\dot{\rho}_t = \frac{i}{\hbar}[\rho_t,\tilde{H}] +\sum_{j,l,{\bf
\delta},{\bf \delta}'}\frac{\Gamma_{j, l}^{{\bf \delta},{\bf
\delta}'}}{2\hbar}\left([b_{j{\bf \delta}},\rho_t b_{l{\bf
\delta'}}] + [b_{l{\bf \delta}'},\rho_t b_{j{\bf \delta}}]\right),\nonumber\\
\label{lindblad}
\end{eqnarray}
with $b_{j{\bf \delta}}=c_{j+{\bf \delta}}^{\dagger}c_{j}$. The
effective system Hamiltonian
\begin{equation} \tilde{H}=-\tilde{J}\sum_{
<i,j>}c_{i}^{\dagger}c_{j}+\tfrac{1}{2}\sum_{i,j}\tilde{V}_{ij}c_{i}^{\dagger}c_{j}^{\dagger}c_{j}c_{i},\label{eq:Heff}\end{equation}
is of the extended Hubbard type, valid for
$\tilde{J},\tilde{V}_{ij},E_p<\Delta$. For $E_p>\Delta$, Eq.(3)
should be derived via a multi-band approach. \emph{Coherent} hopping
of the dressed particles is described by \[\tilde{J}=J\langle\langle
X_{i}^{\dagger}X_{j}\rangle\rangle\equiv J\exp(-S_{T}),\] where
\[S_{T}=\sum_{q}\left(\frac{M_{ q}}{\hbar\omega_{ q}}\right)^{2}[1-\cos({\bf q}{\bf
a})](2n_{ q}(T)+1)\] characterizes the strength of the
particle-phonon interactions, and $n_{q}(T)$ is the thermal
occupation at temperature $T$ \cite{Mahan}.

The dissipative term in Lindblad form in Eq.~\eqref{lindblad}
corresponds to thermally activated \emph{incoherent} hopping with
rates $\Gamma_{j,l}^{{\bf \delta},{\bf \delta}'}$, which can be made negligible for the energies of interest $k_{B}T\ll\min(\Delta,E_p,k_B T_{C})$, see Refs.~\cite{Alexandrov,Ortner}. Corrections to Eq.~\eqref{eq:Heff} proportional to $J^2$ are small relative to
$\tilde{H}$
provided $J\ll E_{p}$~\cite{Alexandrov} (also $J\ll\hbar \omega_{D}$  in 1D~\cite{Ortner,Datta}).
%
%The dissipative term in Lindblad form in Eq.~\eqref{lindblad}
%corresponds to thermally activated \emph{incoherent} hopping with
%rates $\Gamma_{j,l}^{{\bf \delta},{\bf \delta}'}$. In 1D, the latter
%are small compared to $\tilde{J}$ for $S_T\ll 1$ and $S_T\gg 1$,
%provided $J \bar{V}_{q=0}^2 k_B T/[(\hbar \omega_D)^4 \sqrt{r_d}]\ll
%1$ and $k_B T/E_p\ll 1$, respectively \cite{Ortner,Alexandrov}, and
%in particular they are negligible for the energies of interest
%$k_{B}T\ll\min(\Delta,E_p,k_B T_{C})$.
Thus, in the parameter regime of interest
the dynamics of the dressed particles is described by the extended
Hubbard Hamiltonian $\tilde{H}$.  In the following, we verify the existence of this parameter regime and we calculate the effective Hubbard parameters from the microscopic model for the 1D configuration of Fig.~\ref{figs:fig1}(a), where extra-particles are polar molecules of a different species. An analogous calculation for the configuration of Fig.~\ref{figs:fig1}(b) is reported in Ref.~\cite{Pupillo08}.

\begin{center}
\begin{figure}%[h]
\begin{centering}
\includegraphics[width=\columnwidth]{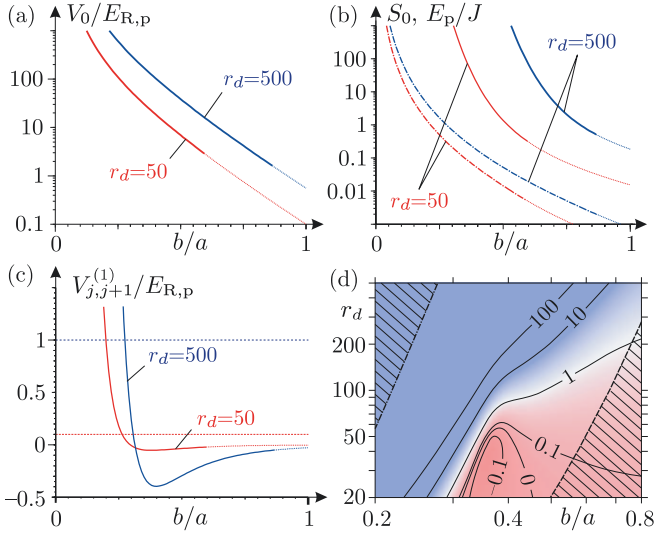}%{fig2_resub}
\par\end{centering}
\caption{\label{figs:fig02} Configuration 1 [Fig.~\ref{figs:fig1}(a)]: Hubbard
parameters for $d_{{\rm p}}/d_{{\rm c}}=0.1$ and $m=m_{{\rm p}}$. (a) Lattice depth $V_{0}$ in units of $E_{{\rm
R,p}}$ vs.~$b/a$ for $r_{d}=50$ and $500$. Thick continuous lines:
tight-binding region $4J<\Delta$. (b) Reduction factor $S_{0}$
(dashed dotted lines) and polaron shift $E_{p}/J$ (solid lines), for
$4J<\Delta$. (c) Continuous lines: phonon-mediated interactions
$V_{j,j+1}^{(1)}$. Horizontal (dashed) lines: $V_{j,j+1}$. (d)
Contour plot of $\tilde{V}_{j,j+1}/2\tilde{J}$ (solid lines) as a
function of $b/a$ and $r_{d}$. A single-band Hubbard model is valid
left of the dashed region ($4 \tilde{J},\tilde{V}_{ij}<\Delta$), and
right of the black region ($E_{p}<\Delta$).}
\end{figure}
\par\end{center}

In the configuration of Fig.~\ref{figs:fig1}(a) molecules of a second
species are trapped in a tube at a distance $b$ from the crystal
tube under 1D trapping conditions. For crystal molecules fixed at
the equilibrium positions with lattice spacing $a$, the extra
particles feel a periodic potential \[V_{{\rm cp}}(x)=d_{{\rm
c}}d_{{\rm p}}\sum_{j}\left[b^{2}+(x-ja)^{2}\right]^{-3/2},\] which
determines the bandstructure. The lattice depth \[V_{0}\equiv
V_{{\rm cp}}(a/2)-V_{{\rm cp}}(0)\sim r_d \frac{d_{\rm p}}{d_{\rm
c}}\frac{m_{\rm p}}{m}\frac{e^{-3 b/a} }{(b/a)^3}E_{\rm
R,p}\] is shown in Fig.~\ref{figs:fig02}(a) as a function of $b/a$,
where the thick solid lines indicate the parameter regime
$4J<\Delta$, and $E_{{\rm R,p}}=\hbar^{2}\pi^{2}/2m_{{\rm p}}a^{2}$.
The potential is comb-like for $b/a<1/4$, since the particles
resolve the individual molecules forming the crystal, while it is
sinusoidal for $b/a\gtrsim1/4$. The strong dipole-dipole repulsion
between the extra particles acts as an effective hard-core
constraint \cite{BuchlerNature07}. We find that for $4J<\Delta$ and
$d_{{\rm p}}\ll d_{{\rm c}}$ the bare off-site interactions satisfy
\[V_{ij}\sim d_{{\rm p}}^{2}/(a|i-j|)^{3}<\Delta,\] which justifies a
single-band approximation for the dynamics of the extra-particles in
the static potential.

The particle-phonon coupling is
\[ M_{q}=\frac{d_{{\rm c}}d_{{\rm p}}}{a b}\left(\frac{2\hbar}{N m_{{\rm
c}}\omega_q}\right)^{1/2}q^2\mathcal{K}_{1}(b|q|)\beta_q\] with
$\mathcal{K}_{1}$ the modified Bessel function of the second kind,
and $M_{q}\sim\sqrt{q}$ for $q\rightarrow0$. In the regime of interest $b/a<1$ where the single-band approximation is valid ($4J,V_{ij}<\Delta$), we find that $M_{q}$ is
peaked at large $q\sim\pi/a$, so that the main contribution to the
integrals in the definition of $S_{T}$ and $E_{p}$ is indeed
dominated by large frequencies $\hbar\omega_{q}>J$. Together with the separation of time-scales $J,V_{ij}\ll \hbar \omega_D$, this is consistent with the picture of the system's dynamics as given by particles dressed by {\em fast} (optical) phonons, as discussed above. We notice that this so-called {\em anti-adiabatic} regime is generally hard to achieve in cold atomic setups \cite{Albus}.

 A plot of
$S_{0}$ as a function of $b/a$ is shown in Fig.~\ref{figs:fig02}(b).
We find the scaling \[S_{0}\propto\sqrt{r_{d}}(d_{{\rm p}}/d_{{\rm
c}})^{2},\] and within the regime of validity of the single band
approximation, $S_{0}$ can be tuned from $S_{0}\ll1$ ($\tilde{J}\sim
J$) to $S_{0}\gg1$ ($\tilde{J}\ll J$) corresponding to the large and
small polaron limit, respectively. The polaron shift $E_{p}$ generally exceeds the bare hopping
rate $J$, and in particular, $E_{p}\gg J$ for $S_{0}\gtrsim1$, see
Fig.~\ref{figs:fig02}(b).
Together with the condition $\hbar\omega_{D}\gg J$ this ensures that
the corrections to Eq.\eqref{eq:Heff} which are proportional to $J^2$
are indeed small, and thus Eq.\eqref{eq:Heff} fully accounts for the coherent dynamics of the dressed particles.

The extended Hubbard model corresponding to the configuration of Fig~\ref{figs:fig1}(a) is characterized by tunable off-site interactions, which are a combination of the direct dipole-dipole interactions between the extra-particles and of the phonon-mediated interactions $V_{i,j}^{(1)}$. For  $b/a\lesssim1/4$ we find that the interactions $V_{i,j}^{(1)}$ decay slowly with the inter-particle distance as  $\sim1/|i-j|^{2}$, and
are thus long-ranged. The sign of $V_{i,j}^{(1)}$ is a function of the ratio $b/a$. Thus, depending on $b/a$ the phonon-mediated interactions can enhance or
reduce the direct dipole-dipole repulsion of the extra particles. As an example, Fig.~\ref{figs:fig02}(c) shows that the sign of the term $V_{j,j+1}^{(1)}$ alternates between attractive and repulsive as a function of $b/a$, and that for small enough $b/a$ the phonon-mediated interactions can become larger than the direct dipole-dipole interactions.

The effective Hubbard parameters
$\tilde{V}_{j,j+1}$ and $\tilde{J}$ are summarized in
Fig.~\ref{figs:fig02}(d), which is a contour plot of
$\tilde{V}_{j,j+1}/2\tilde{J}$ as a function of $r_{d}$ and $b/a$.
The ratio $\tilde{V}_{j,j+1}/2\tilde{J}$ increases by decreasing
$b/a$ or increasing $r_{d}$, and can be much larger than one.
This appearance of strong off-site interactions in the effective dynamics is a necessary ingredient for the realization of a variety of new quantum phases \cite{Opt1,Opt2,Opt3,Opt4,Opt5}. As an
example of the possible quantum phases that can be realized in this
setup, at half filling, and considering nearest-neighbor interactions only, the particles in the configuration described
above undergo a transition from a (Luttinger) liquid
($\tilde{V}_{i,i+1}<2\tilde{J}$) to a charge-density-wave
($\tilde{V}_{i,i+1}>2\tilde{J}$) as a function of $b/a$ and $r_d$. Figure~\ref{figs:fig02}(d) shows that the parameter regime $\tilde{V}_{i,i+1}\approx 2\tilde{J}$, see Ref.~\cite{Hirsch82}, where this transition occurs can be satisfied for various choices of $r_d$ and $b/a$, e.g. for $r_d=100$ and $b/a \approx 0.5$.

\subsection{Three-body interactions}\label{sec:sec3Body}

As discussed in Sect.~\ref{sec:secIntro2}, it is of interest to design systems where effective many-body interactions dominate over the two-body interactions, and determine the properties of the groundstate. We here describe how
an effective low energy interaction potential $V_{\rm eff}$ of the
form Eq.~(\ref{effint1}) can be derived in the
Born-Oppenheimer approximation for $^{1}\Sigma$ polar molecules interacting
via dipole-dipole interactions by dressing low lying rotational
states of each molecule with external static and microwave fields,
in analogy to the discussion of
Sect.~\ref{Sec:TwoMolecules}.

\begin{figure}[htb]
\begin{center}
\includegraphics[width=\columnwidth]{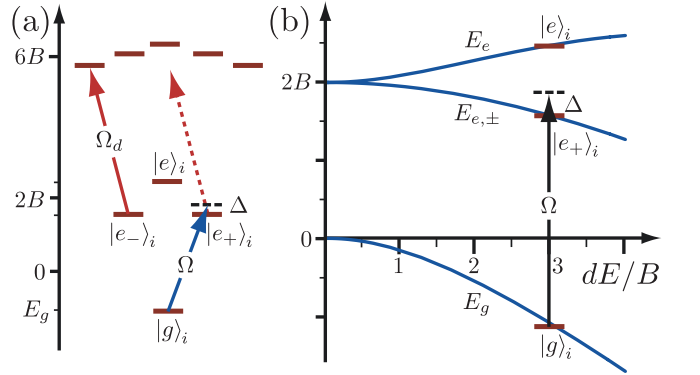}%{fig0-2.eps}
\end{center}
\caption{Spectrum of a polar molecule. (a) Level structure
for $E d/B = 3$: the circular polarized microwave field couples the
ground state $|g\rangle$ with the excited state $|e_{+}\rangle$ with
Rabi frequency $\Omega/\hbar$ and detuning $\Delta$. The excited
state $|e_{+}\rangle$ is characterized by a finite angular momentum
$J_{z} |e_{+} \rangle= |e_{+}\rangle$. Applying a second microwave
field with opposite polarization (red arrow) allows us to lift the
degeneracy in the first excited manifold by resonantly couple the
state $|e_{-}\rangle$ to the next manifold. (b) Internal
excitation energies for a single polar molecule in a static electric
field ${\bf E}= E {\bf e}_{z}$.  } \label{fig0}
\end{figure}

We here focus on a setup with a static electric field ${\bf E} = E
{\bf e}_{z}$ along the $z$-axis, see Fig.~\ref{fig0},   where the
two states  $|g\rangle_{i}\equiv |\phi_{0,0}\rangle_i$ and
$|e_{+}\rangle_{i}\equiv |\phi_{1,+1}\rangle_i$ with energies
$E_{g}$ and $E_{e,\pm}$ are coupled by a circularly polarized
microwave field propagating along the $z$-axis. The microwave
transition is characterized by the (blue) detuning $\Delta>0$ and
the Rabi frequency  $\Omega/\hbar$. While the following discussion can be readily generalized to include the degenerate case \cite{BuchlerNature07}, here we assume that the degeneracy of
the states $|e_{-}\rangle\equiv |\phi_{1,-1}\rangle_i$ and
$|e_{+}\rangle$ is lifted, e.g., by an additional microwave field
coupling the state $|e_{-}\rangle$ near-resonantly to the next state
manifold, see Fig.~\ref{fig0}.
Then, the internal structure of a single polar molecule reduces to a
two-level system and is described as a spin-$1/2$ particle via the
identification of the state $|g\rangle_{i}$ ($|e_{+}\rangle_{i}$) as
eigenstate of the spin operator $S^{z}_{i}$ with positive (negative)
eigenvalue. In the rotating frame and applying the rotating wave
approximation, the Hamiltonian describing the internal dynamics of
the polar molecule reduces to
\begin{equation}
  H^{(i)}_{ 0} = \frac{1}{2} \left( \begin{array}{cc}
    \Delta & \Omega \\
     \Omega & - \Delta
  \end{array} \right)
  = {\bf h} {\bf S}_{i}
\end{equation}
with the effective magnetic field ${\bf h}= (\Omega, 0, \Delta)$ and
the spin operator ${\bf S}_{i}= (S^{x}_{i},S^{y}_{i},S^{z}_{i})$.
The eigenstates of this Hamiltonian are denoted as $|+\rangle_{i} =
\alpha |g\rangle_{i} + \beta |e_{+} \rangle_{i}$ and $|-\rangle_{i}=
- \beta |g \rangle_{i} + \alpha |e_{+} \rangle_{i} $ with energies
$\pm \sqrt{\Delta^2 + \Omega^2}/2$.

For distances $|{\bf r}_{ij}|\gg (D/B)^{1/3}$ with $D= |\langle g|
{\bf d}_{i} |e_{+} \rangle|^2$ and ${\bf d}_{i}$ the dipole
operator, the dipole-dipole interaction Eq.~\eqref{eq:eqDipDip1}
between two polar molecules can be mapped onto the effective spin
interaction Hamiltonian $H_{\rs d} = H^{\rs int}_{\rs d} + H^{\rs
shift}_{\rs d}$. The first term describes an effective spin-spin
interaction
\begin{equation}
 H^{\rs int}_{\rs d}= -\frac{1}{2} \sum_{i \neq j} D \nu({\bf r}_{i j})
\: \left[S_{i}^{x} S_{j}^{x} + S_{i}^{y} S_{j}^{y} -\eta_{-}^2
S_{i}^{z} S_{j}^{z}\right],
\end{equation}
where $\eta_{\pm}= \eta_{g}\pm\eta_{e}$ is determined by the induced
dipole moments $\eta_{g}= \partial_{E} E_{g}/\sqrt{D}$  and
$\eta_{e}= \partial_{E} E_{e,+}/\sqrt{D}$. The anisotropic behavior
of the dipole-dipole interaction is accounted for by $\nu({\bf r}) =
(1-3 \cos^2\vartheta)/r^3$ with $\vartheta$ the angle between ${\bf r}$
and the $z$-axis.  In addition, the asymmetry of the induced dipole
moments gives rise to a position dependent renormalization of the
effective magnetic field and a energy shift
\begin{equation}
 H_{\rs d}^{\rs shift} = \frac{1}{2}\sum_{i\neq j}  D \nu({\bf
r}_{ij})\left[\frac{ \eta_{-}\eta_{+}}{2} S_{i}^{z} +\frac{
\eta_{+}^2}{4} \right]. \end{equation}

\begin{figure}[htb]
\begin{center}
\includegraphics[width=\columnwidth]{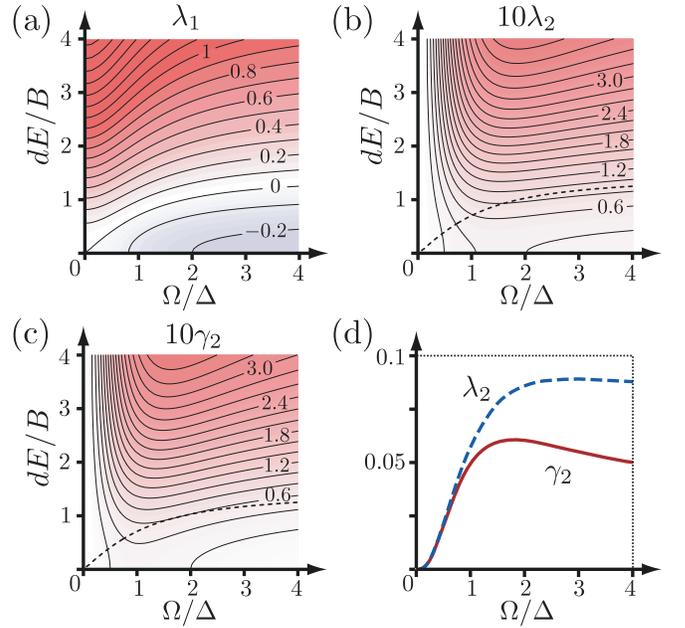}%{fig1-3.eps}
\end{center}
\caption{Parameters of the effective interaction
potential. (a)-(c): Strength of the interaction parameters
$\lambda_{1}$, $\lambda_{2}$, and $\gamma_{2}$ as a function of the
external fields $E d/B$ and $\Omega/\Delta$. The leading
dipole-dipole interaction vanishes for $\lambda_{1}=0$ [dashed line
in (b) and (c)], and the second order contributions dominate
the interaction. (d) Strength of $\lambda_{2}$ (dashed line) and
$\gamma_{2}$ (solid line) along the line in parameter space with
$\lambda_{1}=0$.} \label{fig1}
\end{figure}

Within the Born-Oppenheimer approximation, an analytic expression
for the effective interaction $V_{\rm eff}(\{ {\bf r}_{i}\})$
between two polar molecules each prepared in the state
$|+\rangle_{i}$ can be derived in
second-order perturbation theory in the dipole-dipole interaction
$V_{\rm dd}(\bf r)/{\bf h}$ as
\begin{eqnarray}
V_{\rm eff}(\{ {\bf r}_i\})=E^{(1)}(\{{\bf r}_{i}\})\ +
E^{(2)}(\{{\bf r}_{i}\})\,
\end{eqnarray}
where $D/(a^{3} |{\bf h}|)= (R_{0}/a)^3 \ll 1$ is the (small) parameter
controlling the perturbative expansion, $a$ is the characteristic
length scale of the interparticle separation and $R_{0}=(D/
\sqrt{\Delta^2+\Omega^2})^{1/3}$ is a Condon point, analogous to that discussed in Sect.~\ref{sec:secAC}. The energy shift
\begin{equation}
E^{(1)}(\{{\bf r}_{i}\})\! =\! \frac{1}{2} \left[ \left(\alpha^2
\eta_{g} \!+\! \beta^{2} \eta_{e}\right)^2\! -\!  \alpha^2 \beta^2
\right] \sum_{i \neq j} D \nu({\bf r}_{ij}), \label{firstorder}
\end{equation}
gives rise to a dipole-dipole interaction between the particles,
while the term
\begin{eqnarray}
E^{(2)}\left(\{ {\bf r}_{i}\}\right)& = & \sum_{k\neq i, k\neq j}
\frac{\left| M \right|^2}{\sqrt{\Delta^2+ \Omega^2}} D^2
\nu\left({\bf r}_{i k}\right) \nu\left({\bf r}_{j k}\right)\nonumber\\
& & \hspace{-10pt}+ \sum_{i< j} \frac{\left| N \right|^2}{2
\sqrt{\Delta^2+\Omega^2}}
 \left[D\nu\left({\bf r}_{i j}\right)\right]^2. \label{secondorder}
\end{eqnarray}
corresponds to a correction to the two-particle interaction
potential and an additional three-body interaction.  The matrix
elements $M$ and $N$ take the form
\begin{eqnarray}
  M & = & \alpha \beta \left[ \left(\alpha^2 \eta_{g} + \beta^2
\eta_{e}\right)\left(\eta_{e} - \eta_{g}\right) - (\alpha^2- \beta^2
)/2\right], \nonumber \\ N & = & \alpha^2 \beta^2 \left[
\left(\eta_{e} - \eta_{g} \right)^2 + 1 \right].  \nonumber
\end{eqnarray}
Therefore, the effective interaction potential $V_{\rs eff}$ up to
second order in $(R_{0}/a)^{3}$ reduces to the form in
Eq.~(\ref{effint1}) with the two-particle interaction potential
\begin{equation}
 V({\bf r}) = \lambda_{1} D \: \nu\left({\bf r}\right) + \lambda_{2}
D R_{0}^3 \; \left[\nu\left({\bf r}\right)\right]^2, \label{twobody}
\end{equation}
and the three-body interaction
\begin{eqnarray}
 W\left({\bf r}_{1},{\bf r}_{2},{\bf r}_{3}\right) & =& \gamma_{2} R_{0}^3 D
\label{threebody}
 \left[ \nu({\bf r}_{12})\nu({\bf r}_{13}) \right. \\
& & \hspace{26pt}\left.+ \nu({\bf r}_{12})\nu({\bf r}_{23}) +
\nu({\bf r}_{13})\nu({\bf r}_{23}) \right]. \nonumber
\end{eqnarray}
The dimensionless coupling parameters are $ \lambda_{1}  =
\left(\alpha^2 \eta_{g} + \beta^{2} \eta_{e}\right)^2 - \alpha^2
\beta^2$, $\lambda_{2} = 2 |M|^2 + |N|^2/2$, and $\gamma_{2} = 2
|M|^2$.  These parameters can be tuned via the strength of the
electric field $Ed/B$ and the ratio between the Rabi frequency and
the detuning, $\Omega/\Delta$, see Fig.~\ref{fig1}.  Of special
interest are the values of the external fields where
the leading two-particle interaction vanishes, i.e. $\lambda_{1}=0$. Then, the interaction is dominated by the second order
contribution with $\lambda_{2}$ and $\gamma_{2}$, which includes the
three-body interaction, see Fig.~\ref{fig1}(d), while a small deviation
away from the line $\lambda_{1}=0$ allows us to change the character
of the two-particle interaction. Note, that a $n$-body interaction
term  ($n\geq 4$) appears in $(n-1)\mbox{-th}$ order perturbation
theory in the small parameter $(R_{0}/a)^3$. Therefore, the
contribution of these terms is suppressed
%by the factor $(R_{0}/a)^{3(n-2)}$
and can be safely ignored.

The perturbative expansion requires that $a \gg R_{0}$. We notice
that particles can be confined to interparticle distances larger
than $R_0$ by combining {\em repulsive} dipole-dipole interactions
with a strong (optical) transverse confinement $\omega_{\perp}$, in
analogy to Sect.~\ref{sec:secDC}. Then, for a repulsive
two-particle potential with $\lambda_{1}\gtrsim - \lambda_{2}
(R_{0}/a)^3$ and for $\hbar \omega_{\perp} > D/R_{0}^3$,
two-particles reach distances $|{\bf r}_{i}- {\bf r}_{j}|<R_{0}$ at
an exponentially small rate $\Gamma \sim (\hbar/m a^2) \exp(- 2
S_{\rs E}/\hbar)$, with $S_{\rs E}/\hbar \sim \sqrt{D m/R_{0}
\hbar^2}$. This exponential suppression ensures the stability of the
collisional setup for the duration of an experiment.

The low-energy many-body theory now follows by combining the kinetic
energy of the polar molecules with the  effective interaction
$V_{\rs eff}$ within the Born-Oppenheimer approximation  and the
external trapping potentials $V_{\rs T}$
\begin{equation}
 H= \sum_{i }\frac{{\bf p}^{2}_{i}}{2 m} + V_{\rs eff}\left(\{ {\bf r}_{i}\}\right)
 + \sum_{i}  V_{\rs T}({\bf r}_{i}).
\label{lowenergyhamiltonian}
\end{equation}
This Hamiltonian is independent of the statistics of the particles
and therefore it is valid for bosonic and fermionic polar molecules.

Applying an optical lattice provides a periodic structure for the
polar molecules described by the Hamiltonian
Eq.~\eqref{lowenergyhamiltonian}. In the limit of a deep lattice, a
standard expansion of the field operators $\psi^{\dag}({\bf r})=
\sum_{i} w({\bf r}-{\bf R}_{i}) b_{i}^{\dag}$ in the
second-quantized expression of Eq.~\eqref{lowenergyhamiltonian} in
terms of lowest-band Wannier functions $w({\bf r})$ and particle
creation operators  $b_{i}^{\dag}$ \cite{jaksch98} leads to the
realization of the Hubbard model of Eq.~\eqref{Hubbard},
characterized by strong nearest-neighbor interactions
\cite{BuchlerNature07}. We notice that the particles are
treated as hard-core because of the constraint $a\gg R_0$.
The interaction parameters $U_{i j}$ and $V_{i j k}$ in
Eq.~\eqref{Hubbard} derive from the effective interaction
$V_{\rs}\left( \{ {\bf r}_{i}\}\right)$, and in the limit of
well-localized Wannier functions reduce to
\begin{eqnarray}
U_{i j} & =& U_{0} \frac{ a^3}{|{\bf R}_{i}-{\bf R}_{j}|^3}+
U_{1}\frac{ a^6 }{|{\bf R}_{i}-{\bf R}_{j}|^6},
\end{eqnarray}
and
\begin{eqnarray}
 W_{i j k} & = & W_{0} \left[\frac{a^6}{|{\bf R}_{i}-{\bf R}_{j}|^3|{\bf
R}_{i}-{\bf R}_{k}|^3}+ perm\right], \label{threebodystrength}
\end{eqnarray}
respectively, with $U_{0} = \lambda_{1} D/a^3$, $U_{1}= \lambda_{2}
D R_{0}^3/a^6$, and $W_{0} = \gamma_{2} D R_{0}^3/a^6$. The dominant
contributions and strengths of the three-body terms in different
lattice geometries are shown in Fig.~\ref{figs:fig1000}(b). For  ${\rm Li Cs}$
with a permanent dipole moment $d=6.3 {\rm Debye}$ trapped in an
optical lattice with spacing $a\approx 500{\rm nm}$, the leading
dipole-dipole interaction can give rise to very strong
nearest-neighbor interactions with $U_{0} \sim 55 E_{\rs kin}$, and
$E_{\rs kin}=\hbar^2/m a^2$. On the other hand, tuning the
parameters via the external fields to $\lambda_{1}=0$ the
characteristic energy scale for the three-body interaction becomes
$W_{0} \approx (R_{0}/a)^3 E_{\rs kin}$. Then, controlling the
hopping energy $J$ via the strength of the optical lattice allows to
enter the regime with dominant three-body interactions.
For particles been bosons, an analytic calculation  has suggested that the ground-state phase diagram of Eq.~\eqref{Hubbard} with $U_{ij}=0$ in 1D is characterized by the presence of valence bond states at specific rational fillings of the lattice, charge-density waves and superfluid phases
\cite{BuchlerNature07}.

\subsection{Lattice Spin models}

Cold gases of polar molecules allow to construct in a natural way a
{\em complete toolbox} for any permutation symmetric two spin-$1/2$
(qubit) interaction, using techniques of interaction engineering similar to those discussed in the previous sections.

The system we have in mind is comprised of heteronuclear molecules with $^{2}
\Sigma_{1/2}$ ground electronic states, corresponding for example to
alkaline-earth monohalogenides with a single electron outside a
closed
shell. We adopt a model molecule % (see Fig.~\ref{fig:2})
where the rotational excitations are described by the Hamiltonian
\begin{eqnarray}
H_{\rm m} =B \bf N^2 + \gamma \bf N\cdot \bf S,
\end{eqnarray}
with $\bf N$ the dimensionless orbital angular momentum of the
nuclei, and $\bf S$ the dimensionless electronic spin (assumed to be
$S=1/2$ in the following). Here $B$ denotes the rotational constant
and $\gamma$ is the spin-rotation coupling constant, where a typical
$B$ is a few tens of GHz, and $\gamma$ in the hundred MHz regime.
The coupled basis of a single molecule $i$ corresponding to the
eigenbasis of $H_{\rm m}^i$ is $\{\ket{N_i,S_i,J_i;M_{J_i}}\}$ where
${\bf
  J}_i={\bf N}_i+{\bf S}_i$ with eigenvalues
$E(N=0,1/2,1/2)=0,E(1,1/2,1/2)=2B-\gamma$, and
$E(1,1/2,3/2)=2B+\gamma/2$.

The Hamiltonian describing the internal and external dynamics of a
pair of molecules trapped in wells of an optical lattice is denoted
by $H=H_{\rm in}+H_{\rm ex}$. The interaction describing the
internal degrees of freedom is $H_{\rm in}=H_{\rm dd}+\sum_{i=1}^2
H_{\rm m}^i$, where $H_{\rm dd}$ is the dipole-dipole interaction.
The Hamiltonian describing the
external, or motional, degrees of freedom is $H_{\rm
ex}=\sum_{i=1}^2 {\bf P}_i^2/(2m)+ V_i({\bf x}_i-\bar{{\bf x}}_i)$,
where ${\bf P}_i$ is the momentum of molecule $i$ with mass $m$, and
the potential generated by the optical lattice
$V_i(\bf{x}-\bar{\bf{x}}_i)$ describes an external confinement of
molecule $i$ about a local minimum $\bar{\bf{x}}_i$ with $1$D rms
width $z_0$. We assume isotropic traps that are approximately
harmonic near the trap minimum with a vibrational spacing
$\hbar\omega_{\rm
  osc}$.  Furthermore, we assume that the molecules can be prepared in
the motional ground state of each local potential using dissipative
electromagnetic pumping~\cite{DeMille:05}.  It is convenient to
define the quantization axis $\hat{z}$ along the axis connecting the
two molecules, $\bar{\bf{x}}_2-\bar{\bf{x}}_1=\Delta z \hat{z}$ with
$\Delta z$ corresponding to a multiple of the lattice spacing.

The ground subspace of each molecule is isomorphic to a spin $1/2$
particle.  Our goal is to obtain an effective spin-spin interaction
between two neighboring molecules.  Static spin-spin interactions
due to spin-rotation and dipole-dipole couplings do exist but are
very small in our model: $H_{\rm vdW}(r) = -(d^4/2Br^6)
\left[1+\left(\gamma/4B\right)^2\left(1+4{\bf
      S}_1\cdot{\bf S}_2/3-2S_1^zS_2^z \right) \right] $.  The first
term is the familiar van-der-Waals $1/r^6$ interaction, while the
spin dependent piece is strongly suppressed as $\gamma / 4B
\approx10^{-3} \ll 1$.  However,
dipole-dipole coupled excited states can be dynamically mixed using a microwave field.

\begin{figure*}
  \begin{center}
   \includegraphics[width=\textwidth]{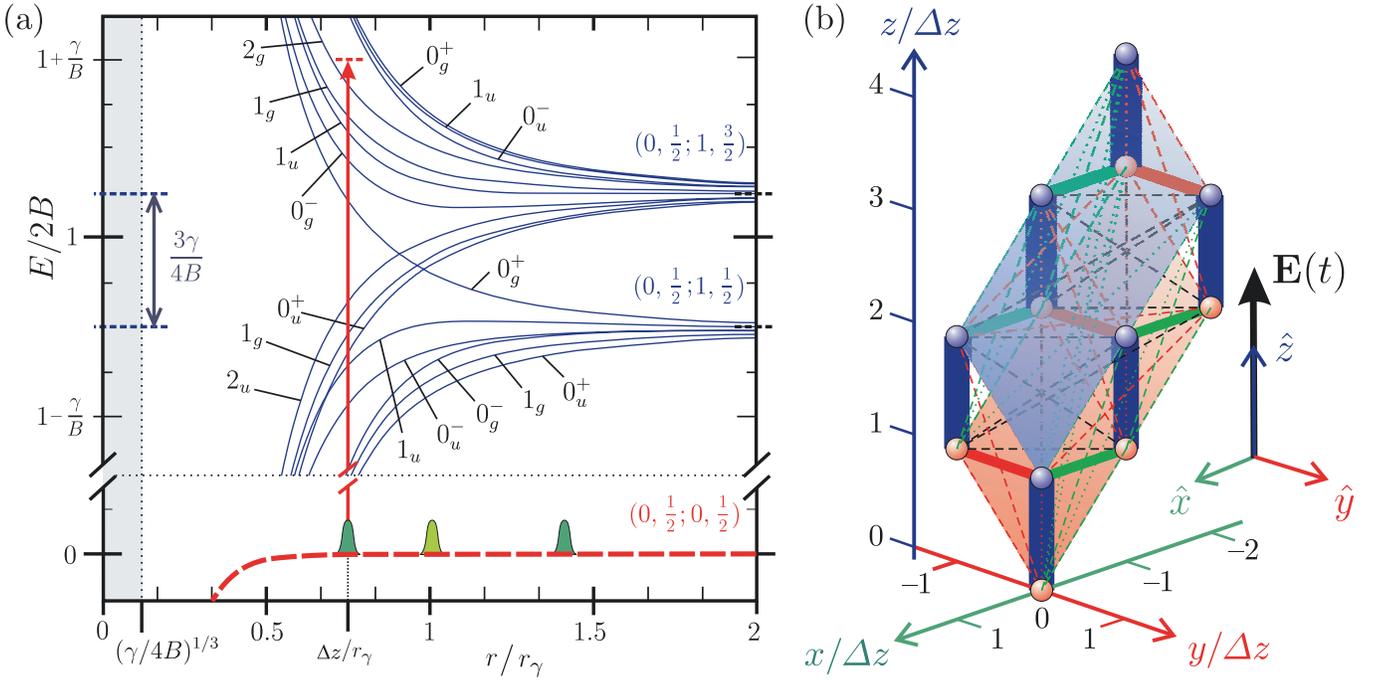}%{Rev23_Fig2.eps}
    \caption{\label{fig:2}(a) Movre-Pichler potentials for
      a pair of molecules as a function of their separation $r$: The
      potentials $E(g_i(r))$ for the $4$ ground-state (dashed lines)
      and the potentials $E(\lambda(r))$ for the first $24$ excited
      states (solid lines). The symmetries $|Y|^\pm_\sigma$ of the
      corresponding excited manifolds are indicated, as are the
      asymptotic manifolds $(N_i,J_i;N_j,J_j)$. (b) Implementation of spin model $H_{\rm spin}^{({\rm II})}$.
    Shown is the spatial configuration of $12$ polar molecules trapped by two parallel triangular
    lattices (indicated by shaded planes) with separation normal to the plane of $\Delta z/\sqrt{3}$ and
    in plane relative lattice shift of $\Delta z\sqrt{2/3}$.  Nearest neighbors are separated by $b=\Delta z$ and next nearest neighbor couplings are at $\sqrt{2}b$.  The graph vertices represent spins and the edges correspond to pairwise spin couplings.  The edge color indicates the nature of the dominant
   pairwise coupling for that edge (blue$=\sigma^z\sigma^z$, red$=
   \sigma^y\sigma^y$, green$=\sigma^x\sigma^x$, black$=$``other").   For nearest neighbor couplings, the edge width indicates the relative strength of the absolute value of the coupling.
   For this implementation, the nearest neighbor separation is $b=r_{\gamma}$.
   Three fields all polarized along $\hat{z}$ were used to generate the effective spin-spin interaction with frequencies and intensities optimized to approximate the ideal model $H_{\rm spin}^{({\rm II})}$.  The field detunings at the nearest neighbor spacing are:  $ \hbar\omega_1-E(1_g(1/2))=-0.05 \gamma/2, \hbar\omega_2-E(0_g^-(1/2))=0.05 \gamma/2,\hbar\omega_3-E(2_g(3/2))=0.10 \gamma/2$ and the amplitudes are $|\Omega_1|=4|\Omega_2|=|\Omega_3|=0.01 \gamma/\hbar$.  For $\gamma=40 {\rm MHz}$ this generates effective coupling strengths $J_z=-100 {\rm kHz}$ and $J_{\perp}=- 0.4 J_z$.  The magnitude of residual nearest neighbor couplings are less than $0.04|J_z|$ along $x$ and $y$-links and less than $0.003|J_z|$ along $z$-links.  The size of longer range couplings $J_{\rm lr}$ are indicated by edge line style (dashed: $|J_{\rm lr}|<0.01 |J_z|$, dotted:  $|J_{\rm lr}|<10^{-3} |J_z|$).    Treating pairs of spins on $z$-links as a single effective spin in the low energy sector, the model approximates Kitaev's $4$-local Hamiltonian \cite{DKL:03} on a square grid (shown here are one plaquette on the square lattice and a neighbor plaquette on the dual lattice) with an effective coupling strength $J_{\rm eff}=-(J_{\perp}/J_z)^4|J_z|/16\approx 167 {\rm Hz}$.}
  \end{center}
  \end{figure*}

The molecules are assumed to be trapped with a separation $\Delta z\sim
r_\gamma\equiv(2d^2/\gamma)^{1/3}$, where the dipole dipole
interaction is $d^2/r_\gamma^3=\gamma/2$.  In this regime the
rotation of the molecules is strongly coupled to the spin and the
excited states are described by Hunds case (c) states in analogy to
the dipole-dipole coupled excited electronic states of two atoms
with fine-structure.  The ground states are essentially spin
independent.  In the subspace of one rotational quantum
$(N_1+N_2=1)$, there are $24$ eigenstates of $H_{\rm in}$ which are
linear superpositions of two electron spin states and properly
symmetrized rotational states of the two molecules.  There are
several symmetries that reduce $H_{\rm in}$ to block diagonal form.
First, $H_{\rm dd}$, conserves the quantum number $Y=M_N+M_S$ where
$M_N=M_{N_1}+M_{N_2}$ and $M_S=M_{S_1}+M_{S_2}$ are the total
rotational and spin projections along the intermolecular axis.
Second, parity, defined as the interchange of the two molecules
followed by parity though the center of each molecule, is conserved.
The $\sigma=\pm 1$ eigenvalues of parity are conventionally denoted
$g(u)$ for {\it g}erade({\it u}ngerade). Finally, there is a
symmetry associated with reflection $R$ of all electronic and
rotational coordinates through a plane containing the intermolecular
axis.  For $|Y|>0$ all eigenstates are even under $R$ but for states
with zero angular momentum projection there are $\pm 1$ eigenstates
of $R$.  The $16$ distinct eigenvalues correspond to degenerate
subspaces labeled $|Y|_{\sigma}^{\pm}(J)$ with $J$ indicating the
quantum number in the $r\rightarrow \infty$ asymptotic manifold
$(N=0,J=1/2;N=1,J)$. Remarkably, the eigenvalues and eigenstates can
be computed analytically yielding the Movre-Pichler
potentials~\cite{Movre:77} plotted in Fig.~\ref{fig:2}(a).\\

In order to induce strong dipole-dipole coupling we introduce a
microwave field $E({\bf x},t){\bf e}_F$ with a frequency $\omega_F$
and Rabi-frequency $\Omega$ tuned near resonance with the
$N=0\rightarrow N=1$ transition.

The effective Hamiltonian acting on the lowest-energy states is obtained in
second order perturbation theory as
\begin{equation}
  H_{\rm eff}(r)=\sum_{i,f}\sum_{\lambda(r)}\frac{\bra{g_f}H_{\rm mf}\ket{\lambda(r)}\bra{\lambda(r)}H_{\rm mf}\ket{g_i}}{\hbar\omega_F-E(\lambda(r))}\ket{g_f}\bra{g_i},
  \label{effHam1}
\end{equation}
where $\{{\ket{g_i},\ket{g_f}}\}$ are ground states with $N_1=N_2=0$
and $\{\ket{\lambda(r)}\}$ are excited eigenstates of $H_{\rm in}$
with $N_1+N_2=1$ and with excitation energies $\{E(\lambda(r))\}$.
The reduced interaction in the subspace of the spin degrees of
freedom is then obtained by tracing over the motional degrees of
freedom.  For molecules trapped in the ground motional states of
isotropic harmonic wells with rms width $z_0$ the wave function is
separable in center of mass and relative coordinates,
and the effective spin-spin Hamiltonian is $H_{\rm
spin}=\langle H_{\rm  eff}(r)\rangle_{\rm rel}$.

The Hamiltonian in Eq.~\eqref{effHam1} is guaranteed to yield some
entangling interaction for appropriate choice of field parameters
but it is desirable to have a systematic way to design a spin-spin
interaction. The model presented here possesses
sufficient structure to achieve this essentially analytically.  The
effective Hamiltonian on molecules $1$ and $2$ induced by a
microwave field is
\begin{equation}
  H_{\rm eff}(r)=\frac{\hbar|\Omega|}{8}\sum_{\alpha,\beta=0}^3 \sigma^{\alpha}_1A_{\alpha,\beta}(r)\sigma^{\beta}_2,
  \label{effHam2}
\end{equation}
where $\{\sigma^{\alpha}\}_{\alpha=0}^3\equiv \{{\bf
  1},\sigma^x,\sigma^y,\sigma^z\}$ and $A$ is a real symmetric tensor.

Equation~\eqref{effHam2} describes a generic permutation symmetric two
qubit Hamiltonian.  The components $A_{0,s}$ describe a pseudo
magnetic field which acts locally on each spin and the components
$A_{s,t}$ describe two qubit coupling.  The pseudo magnetic field is
zero if the microwave field is linearly polarized but a real
magnetic field could be used to tune local interactions and, given a
large enough gradient, could break the permutation invariance of
$H_{\rm spin}$.

For a given field polarization, tuning the frequency near an excited
state induces a particular spin pattern on the ground states.  These
patterns change as the frequency is tuned though multiple resonances
at a fixed intermolecular separation.  In Table~\ref{tab:2} it is
shown how to simulate the Ising and Heisenberg interactions in this
way.  Using several fields that are sufficiently separated in
frequency, the resulting effective interactions are additive
creating a {\it spin texture} on the ground states.  The anisotropic
spin model $H_{XYZ}=\lambda_x \sigma^x\sigma^x+\lambda_y
\sigma^y\sigma^y+\lambda_z\sigma^z\sigma^z$ can be simulated using
three fields: one polarized along $\hat{z}$ tuned to $0_u^+(3/2)$,
one polarized along $\hat{y}$ tuned to $0_g^-(3/2)$ and one
polarized along $\hat{y}$ tuned to $0_g^+(1/2)$.  The strengths
$\lambda_j$ can be tuned by adjusting the Rabi frequencies and
detunings of the three fields.  Using an external magnetic field and
six microwave fields with, for example, frequencies and
polarizations corresponding to the last six spin patterns in
Table~\ref{tab:2}, arbitrary permutation symmetric two qubit
interaction are possible.

\begin{table}
  \caption{\label{tab:2}  Some spin patterns that result from Eq.~\eqref{effHam2}.  The field polarization is given with respect to the intermolecular axis $\hat{z}$ and the frequency $\omega_F$ is chosen to be near resonant with the indicated excited state potential at the internuclear separation $\Delta z$.  The sign of the interaction will depend on whether the frequency is tuned above or below resonance.}
  \begin{ruledtabular}
    \begin{tabular}{ccc}
      \hline
      Polarization &\  Resonance &\  Spin pattern\\
      \hline
      $\hat{x}$ &\  $2_g$ &\  $\sigma^z\sigma^z$\\
      \hline
      $\hat{z}$ &\  $0_u^+$ &\  $\vec{\sigma}\cdot\vec{\sigma}$ \\
      \hline
      $\hat{z}$ &\  $0_g^-$ &\  $\sigma^x\sigma^x+\sigma^y\sigma^y-\sigma^z\sigma^z$ \\
      \hline
      $\hat{y}$ &\  $0_g^-$ &\  $\sigma^x\sigma^x-\sigma^y\sigma^y+\sigma^z\sigma^z$ \\
      \hline
      $\hat{y}$ &\  $0_g^+$ &\  $-\sigma^x\sigma^x+\sigma^y\sigma^y+\sigma^z\sigma^z$ \\
      \hline
      $(\hat{y}-\hat{x})/\sqrt{2}$ &\  $0_g^+$ &\  $-\sigma^x\sigma^y-\sigma^y\sigma^x+\sigma^z\sigma^z$ \\
      \hline
      $\cos\xi\hat{x}+\sin\xi\hat{z}$ &\  $1_g$ &\  $\lambda_1(\sigma^x\sigma^z+\sigma^z\sigma^x)+\lambda_2\sigma^z\sigma^z$\\
      &\ &\ $+\lambda_3(\sigma^x\sigma^x+\sigma^y\sigma^y)$ \\
      \hline
      $\cos\xi\hat{y}+\sin\xi\hat{z}$ &\  $1_g$ &\  $\lambda_1(\sigma^y\sigma^z+\sigma^z\sigma^y)+\lambda_2\sigma^z\sigma^z$\\
      &\ &\ $+\lambda_3(\sigma^x\sigma^x+\sigma^y\sigma^y)$ \\
    \end{tabular}
  \end{ruledtabular}
\end{table}

The Kitaev model of Eq.~\eqref{Kit} (Spin model ${\rm II}$) can be obtained in the following way. %is
%likewise obtained using this mechanism.
Consider a system of four
molecules connected by three length $b$ edges forming an orthogonal
triad in space. There are several different microwave field
configurations that can be used to realize the interaction $H_{\rm
spin}^{({\rm II})}$ along the links. One choice is to use two
microwave fields polarized along $\hat{z}$, one tuned near resonance
with a $1_g$ potential and one near a $1_u$ potential.
 A realization of model {\rm II} using
 a different set of $3$ microwave fields is shown in Fig.~\ref{fig:2}(b).
The obtained interaction is close to ideal with small residual
coupling to next nearest neighbors.% as in model ${\rm I}$.

%\end{multicols}
\end{document}